\documentclass[sigconf]{acmart}
\usepackage{array} 
\usepackage[utf8]{inputenc}
\usepackage{graphicx}
\usepackage{dblfloatfix} 
\usepackage{booktabs}
\usepackage{colortbl}

\copyrightyear{2024}
\acmYear{2024}
\setcopyright{rightsretained}
\acmConference[ASSETS '24]{The 26th International ACM SIGACCESS Conference on Computers and Accessibility}{October 27--30, 2024}{St. John's, NL, Canada}
\acmBooktitle{The 26th International ACM SIGACCESS Conference on Computers and Accessibility (ASSETS '24), October 27--30, 2024, St. John's, NL, Canada}
\acmDOI{10.1145/3663548.3675613}
\acmISBN{979-8-4007-0677-6/24/10}

\title{Engaging with Children's Artwork in Mixed \\Visual-Ability Families}

\author{Arnavi Chheda-Kothary}
\affiliation{%
   \department{Paul G. Allen School of Computer Science \& Engineering | DUB Group}
   \institution{University of Washington}
   \city{Seattle}
   \state{WA}
   \country{USA}}
\email{chheda@cs.washington.edu}

\author{Jacob O. Wobbrock}
\affiliation{%
    \department{The Information School | DUB Group}
   \institution{University of Washington}
   \city{Seattle}
   \state{WA}
   \country{USA}}
\email{wobbrock@uw.edu}

\author{Jon E. Froehlich}
\affiliation{%
   \department{Paul G. Allen School of Computer Science \& Engineering | DUB Group}
   \institution{University of Washington}
   \city{Seattle}
   \state{WA}
   \country{USA}}
\email{jonf@cs.uw.edu}

\begin{teaserfigure}
\centering
\includegraphics[width=\textwidth]
{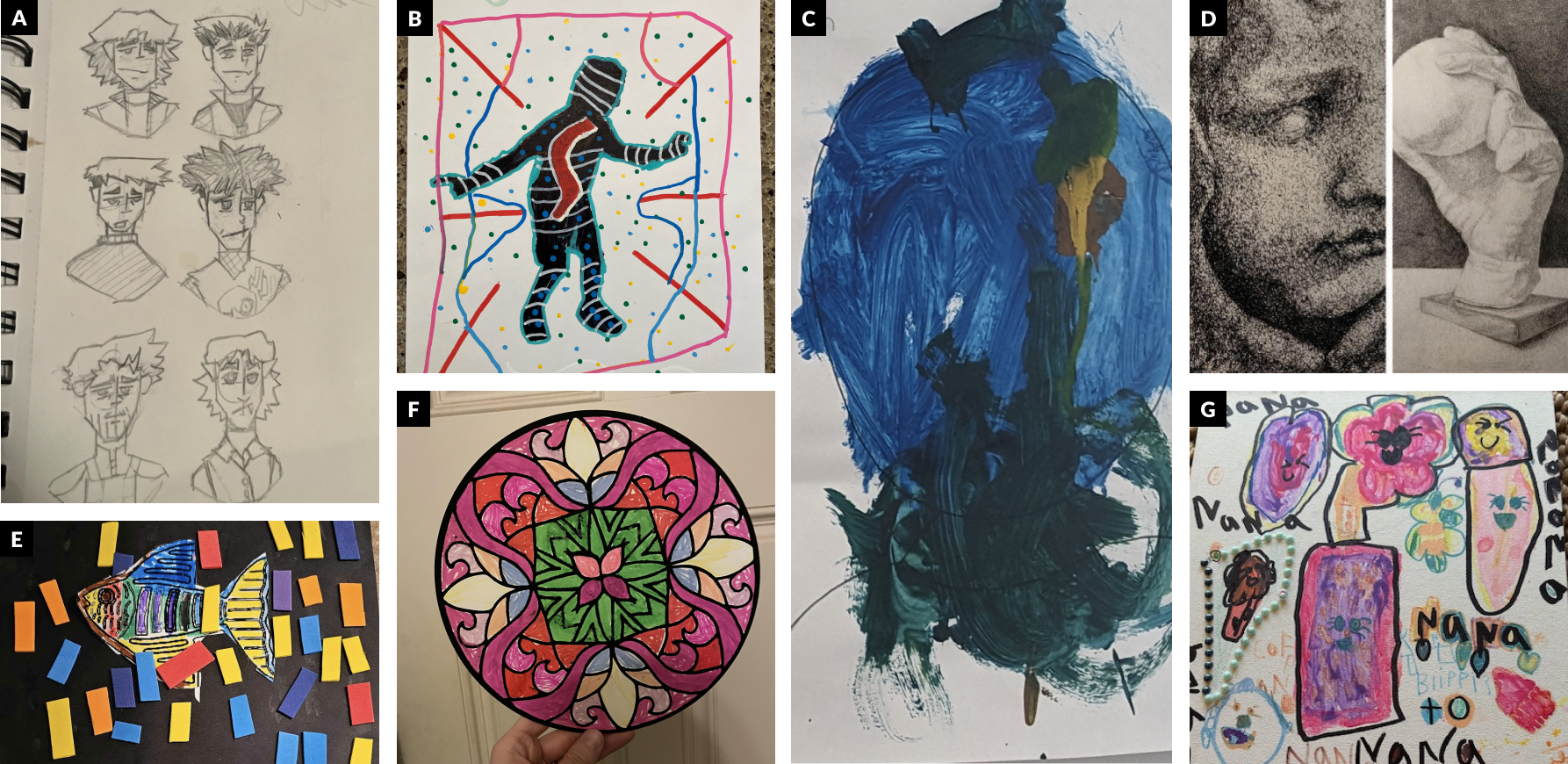}
\Description{A collage of 7 images, labelled A through G, which are all different types of children's artwork. The images are laid out in a grid-like format, with A through D in the top row and E, F, and G in the bottom row. Image A is a page from a sketchbook that shows 6 different characters drawn in pencil, all with variations of spiky hair and different types of confused facial expressions. Image B is a coloring of a black silhouette figure striking a pose with its hands out by its side, and the there are white lines and a prominent red squiggle on the figure. The background has different colored lines and dots. Image C is an abstract painting with large blue and green brushstrokes, and a touch of yellow. Image D consists of two art pieces, both extremely detailed and very well done. The art piece on the left shows part of a young child's face in black and white; the art piece on the right shows a statue of a hand holding a ball between its thumb and index and middle fingers, also in black and white. Image E has a colorful fish on black paper, and different colored rectangles in orange, red, yellow, blue, purple that are surrounding the fish. Image F is a mandala that has a symmetrical design with a green center that contains a pink and purple flower, and shades of pink, red, and yellow on the outside. Image G has a series of different figures and flowers drawn in pink and purple colors, as well as the word "Nana" repeated throughout the image; it has been done by younger children.}
\caption{Children in BLV families created a variety of artwork, including: (A) age 14, at-home character sketches, (B) age 9, school project, (C) age 4, school project, (D) age 17, artwork for her portfolio, (E) age 7, school project, (F) age 6, school project, (G) ages 5 and 7, joint drawing done at home. We have removed identifying information from all artwork in this paper.}
\end{teaserfigure}
\vspace{-0.2em}

\begin{abstract}
We present two studies exploring how blind or low-vision (BLV) family members engage with their sighted children's artwork, strategies to support understanding and interpretation, and the potential role of technology, such as AI, therein. Our first study involved 14 BLV individuals, and the second included five groups of BLV individuals \textit{with} their children. Through semi-structured interviews with AI descriptions of children's artwork and multi-sensory design probes, we found that BLV family members value artwork engagement as a bonding opportunity, preferring the child's storytelling and interpretation over other nonvisual representations. Additionally, despite some inaccuracies, BLV family members felt that AI-generated descriptions could facilitate dialogue with their children and aid self-guided art discovery. We close with specific design considerations for supporting artwork engagement in mixed visual-ability families, including enabling artwork access through various methods, supporting children's corrections of AI output, and distinctions in \textit{context vs. content} and \textit{interpretation vs. description} of children's artwork. 
\end{abstract}

\begin{document}

\maketitle
\section{Introduction}

\begin{quote}
    \textit{When a child produces something, whether it's a piece of artwork or a project, they want to share that joy. So they are looking for affirmation; they are looking for praise...``that's beautiful.'' But how do you express to your children that you're impressed with their drawing while not being able to see the drawing?} --- \textbf{P3}
\end{quote}

Artistic creation is intrinsic to childhood. Through art, children creatively express themselves, impacting happiness, social skills, intellectual development, and confidence \cite{importance_for_early_development_2024, Lynch_2012, Rymanowicz_2021}. To support the joy and creativity that artistic expression fosters in children, it is important that parents and relatives engage with their children’s artwork \cite{how_parents_can_help_2024}. Art-based interactions between parents and children, such as co-creating artwork or participating in art therapy, also enable deeper parent-child closeness \cite{Sakr2017} and positively affect children's perceptions of acceptance \cite{Shamri_2018}. While the importance of artistic creation and art-based interactions between child and parent is well-established, there are open questions about how blind or low-vision (BLV) parents and relatives engage with their sighted children's artwork, such as how they perceive visual art creations, provide encouragement and feedback, and what, if any, tools and techniques they might use to help them explore their children's creative output.

Although recent research and technology advances in artificial intelligence (AI) and digital fabrication (3-D printers, laser cutters) have explored nonvisual access to visual artifacts, including BLV-accessible art creation tools \cite{Luebs_ExpertCraftingPractices}, tactile representations of professional artwork \cite{Butler_MultiSensory, Holloway_MakingSenseOfArt}, and AI-based applications for image descriptions \cite{seeing_ai, be_my_ai_2024}, a gap remains in our understanding of how BLV family members engage with visual creations made by their children. In this work, we explore: (1) How do BLV family members \textit{currently engage} with their children's artwork? (2) What \textit{information} about their children's artwork do BLV family members \textit{want} to have available? (3) How can BLV family members \textit{use technology} to enhance engagement with their children's artwork?

To investigate these questions, we conducted two semi-structured interview studies. In Study 1, we recruited 14 BLV family members and asked questions about current strategies to support understanding and interpretation of their children's artwork, motivations for and barriers to doing so, and their use of technology in aiding comprehension. For Study 2, we recruited a subset of BLV family members from Study 1 with their children; here our focus was on the child, the child's process of creating and sharing art, and the child-adult interaction. Both studies gathered feedback about AI interpretations of their children's artwork using two state-of-the-art tools: \textit{Be My AI} \cite{be_my_ai_2024} and \textit{ChatGPT4} \cite{explore_gpts_2024}. The second study also utilized design probes with two in-person relative-and-child groups to explore non-visual art representations preferences. Interview sessions were recorded, transcribed, and then iteratively coded to reach a set of themes concerning current practices and underlying motivations, desired information about artwork, and responses to actual or potential technology.

Our findings reveal that while fostering an emotional connection is the primary motivation for learning about their children’s artwork, BLV parents also wanted to use their children’s art to monitor developmental progress. For the BLV family member to gain understanding of the artwork, children would begin with a verbal description, which would often include contextual information beyond the visual characteristics of the art, such as where the art was made and what inspired the art. Most BLV members were also curious about additional details beyond their children’s description, but the specific details (\textit{e.g.}, colors, techniques) and the desired access mechanism to interpret those details (\textit{e.g.}, tactile, audio description) varied based on the child's age, the artwork style, and the BLV relatives' art preferences. 

When presented with AI-based descriptions of their children's artwork, most BLV family members found them valuable even if the descriptions contained inaccuracies. Children, however, wanted to correct erroneous AI descriptions and provide additional personal and contextual details that would be imperceptible to the AI (\textit{e.g.,} why an artistic choice was made, the inspiration for the artwork). Finally, two in-person design probes exploring different technological representations of children's artwork revealed BLV family members' and their sighted children's preferences for audio descriptions and tactile representations of child-created art.

In summary, our work contributes both children's and BLV family members' perspectives to advance understanding of: (1) current practices that mixed visual-ability families use to engage with children's artwork, (2) information that BLV family members want to know about their children's artwork, and (3) ways that mixed visual-ability families can use technology, particularly AI, to supplement how they experience and understand children's artwork.

\section{Background and Related Work}
Our work builds on prior research in non-visual representations of artwork and images, the emerging role of AI, and how technology supports mixed-ability interactions.

\subsection{Nonvisual Representation of Art and Images}
There is a rich history of work in human-computer interaction (HCI) investigating nonvisual representations of art and images for blind and low-vision (BLV) people \cite{Morris_RichRepresentations, Singh_FigureA11y, Mack_HighQualityAltText, Edwards_HowAltTextGetsMade, Bellscheidt_AuthoringAltText, WilliamsDeGreef_QualityAltTextInComputing, Kim_DescriptionsForComics, effective_alt_text}. In the digital realm, BLV people primarily interpret images through access technology such as screen readers, which read human- or AI- generated alternative text descriptions (ALT text) \cite{effective_alt_text, Bellscheidt_AuthoringAltText, Morris_RichRepresentations}. Several commercial applications, including \textit{Be My Eyes}~\cite{be_my_eyes_2024} and \textit{Seeing AI}~\cite{seeing_ai}, provide human or AI description services for live scenes as well as images.

Researchers have increasingly explored alternate modalities to represent images and graphics, including touchscreen-based image interactions \cite{Lee_ImageExplorer, Nair_ImageAssist}, tactile representations \cite{Butler_SystematicReviewOfTactile}, electrotactile feedback \cite{Jiang_Electrotactile}, vibrofeedback \cite{Hong_EvaluatingHapticFeedback_ASSETS2017}, sonification \cite{Holloway_Infosonics}, and multimodal systems \cite{Bartolome_ArtMultimodalExploration, Sharif_VoxLens, Stearns_HandSight_TACCESS2017}. Tactile graphics creation techniques include 3D printing \cite{Shi_MagicTouch}, swell touch paper \cite{Holloway_MakingSenseOfArt}, laser cutting \cite{Holloway_MakingSenseOfArt, Butler_MultiSensory}, refreshable tactile displays \cite{dotpad}, and do-it-yourself (DIY) strategies \cite{Zuniga_LowCostDIY}. Touchscreens afford direct spatial input via touch and audio output, but lack tactility. Tactile representations offer high tactility but can be more challenging to produce, and static, rather than dynamic. Multimodal systems can combine input and output modalities, but are often more complex to use and unavailable to the general public. Advances in materials, fabrication, and AI might eventually remedy some of these trade-offs.

A smaller body of work studies artwork interpretation, predominantly in the context of museums or professional artwork. \citet{Asakawa_IndependentMuseumExperience} explore an interactive museum experience for blind people using an ``art appreciation'' mode on a mobile application that provides an auditory description of artwork. ~\citet{Butler_MultiSensory} and ~\citet{Holloway_MakingSenseOfArt} evaluate BLV people's responses to different sensory-based artwork representations including 3-D printed art, laser cuts, and swell paper graphics. ~\citet{Cavazos_MultiModalArtA11y} investigate a multimodal approach to interpret a Vincent van Gogh painting, and \citet{Rector_EyesFree} investigate the use of proxemic audio to experience paintings using different sonification and verbal description strategies. ~\citet{Li_UnderstandingVisualArtsExperiences} conduct a mixed-methods study involving interviews and a survey to understand motivations and practices BLV individuals use to enjoy visual arts. Though this prior work has demonstrated the effectiveness of tactile and detailed audio descriptions for museum art, and motivations to understand professional visual art, a gap remains in identifying practices and the potential for technological involvement when BLV people interpret children's artwork, specifically in family settings. 

\subsection{AI as a Tool for BLV People}
With recent advances in generative AI (GAI), new AI-powered BLV tools have emerged such as \textit{Be My AI}~\cite{be_my_ai_2024}, a GPT-backed \cite{explore_gpts_2024} system available through \textit{Be My Eyes}; \textit{OKO}~\cite{oko}, a navigation application for pedestrian safety that also uses GPT; and AI advancements in screen readers such as JAWS~\cite{jaws_software_2024} for improved picture interpretation.

Research by \citet{gonzalez2024investigating} explores BLV people's usage and trust of AI tools to describe visual content. Through a diary study, they identify ways to improve AI descriptions to increase trustworthiness to BLV people. ~\citet{Huh_GenAssist} leverage different AI models to unlock a more accessible experience for the BLV community when creating and vetting AI-generated images. ~\citet{Glazko_Autoethnography} report on self-usage of GAI tools; a BLV researcher on the team talked through pros and cons of using GAI to write code and vet generated user interface layouts. However, the effects of the latest advances in GAI have not yet been evaluated in the context of mixed visual-ability families. Additionally, there is scant understanding of how these AI tools perform on diverse types of children's drawings.

\subsection{Supporting Mixed-Ability Groups}
A \textit{mixed-ability group} is any group with some members having a disability and others not~\cite{Storer_Branham}. We draw upon prior research investigating technology in mixed-ability spaces for our work in mixed \textit{visual}-ability families. For example, ~\citet{Phutane_2022} investigate tactile materials by teachers of BLV students, highlighting how tactile as a medium can be used for communicating visual concepts in mixed visual-ability settings. Branham and Kane employ semi-structured interviews and field studies as qualitative methods to examine how blind and sighted companions co-create accessible home spaces~\cite{Branham_Kane_HomeSpaces} and how blind employees navigate mixed-ability workplaces~\cite{Branham_Kane_Workplaces}; our studies are methodologically similar, as we interview both the BLV adult and their sighted collaborator (in our work's case, their children) to understand collective perspectives on artwork sharing. Artifacts showing potential for technology in mixed-ability settings include dual interfaces that fuse visual and nonvisual interactions by \citet{savidis1995developing}, a multimodal handwriting-learning environment for BLV students and their teachers by ~\citet{Plimmer_Multimodal}, and a prompter system by ~\citet{Carmien_EndUserProgramming} supporting young adults with cognitive disabilities and their caregivers. Moreover, research in mixed-ability family settings includes work by ~\citet{Storer_Branham}, ~\citet{Park_AllAboutPictures_2023}, and ~\citet{Cassidy_CuddlingUp}, who explore the design of technology for co-reading interactions between BLV parents and their children, approaching the BLV parent's perspective through observational and interview studies like we do in our work. However, prior research has not investigated technology in mixed visual-ability families' interactions around children's artwork.

\section{Study 1: Interviews with BLV Adults}
To explore how blind and low-vision (BLV) family members' and their sighted children engage with child-created art and strategies to support understanding and interpretation, we conducted a semi-structured interview with 14 BLV adult participants (Study 1). We then conducted a complementary study (Study 2) with five BLV family members \textit{and} their children to gain an understanding of the child's perspective, to directly observe responses to AI and tactile interpretations of their work, and to examine cross-generational dynamics. We report on Study 1 below.

\subsection{Participants}
We recruited 14 BLV family members of sighted children, including both parents and non-parent relatives (\textit{e.g.}, grandparents) through email lists, social media posts, and snowball sampling \cite{snowball_sampling}. All family members were 18 years of age or older. Participants were located throughout the United States and had the option of participating virtually through Zoom, over the phone, or in person. Table 1 shows participant demographics and related information.

\begin{table*}[t]
\centering
\LARGE 
\setlength{\extrarowheight}{5pt} 
\resizebox{\textwidth}{!}{ 
\rowcolors{2}{white}{gray!10}
\begin{tabular}{cccp{3.25cm}p{6.5cm}lp{2.5cm}p{2cm}p{3.5cm}}
\toprule 
\rowcolor{white}  
\textbf{PID} & \textbf{Age Range} & \textbf{Gender} & \textbf{Degree of Vision Loss} & \textbf{Additional Vision Loss Details} & \textbf{Age of Children} & \textbf{Relationship to Children} & \textbf{Provided Artwork?} & \textbf{Participated in Study 2?}\\
\midrule
1 & 25-34 & F & Totally Blind/No Usable Vision & Leber's Congenital Amaurosis (LCA), some light perception, since birth & Between 6-17 & Cousin & No & No\\

2 & 45-54 & M & Legally blind & No vision in left eye, some vision in right eye, onset in 2002 & 13 and 14 & Father & Yes & No\\

3 & 45-54  & M & Legally blind & Retinitis pigmentosa in both eyes, onset at age 32 & 8 & Uncle & Yes & No\\

4 & 65+ & F & Totally Blind/No Usable Vision & Some light perception, and can see some color if very close; vision loss developed later in life & 4 & Grandmother & No & No\\

5 & 55-64 & F & Legally Blind & Blind from birth, visual acuity is 20/200 - 20/400 with nystagmus & 17 & Mother & Yes & No\\

6 & 18-24 & M & Totally Blind/No Usable Vision & Leber's Congenital Amaurosis (LCA), some light perception, since birth & 7 & Father & Yes & Yes, over Zoom with 7-year old son\\

7 & 35-44 & M & Totally Blind/No Usable Vision & Legally blind since birth, total blindness at age 18 & 7 and 9 & Father & Yes & Yes, over Zoom with 7-year old daughter\\

8 & 65+ & F & Legally Blind & 5\% vision left in both eyes due to glaucoma and diabetic retinopathy, developed later in life & 5, 7, 9 & Great-Grandmother & Yes & Yes, in-person with all three great-grandchildren\\

9 & 35-44 & F & Legally Blind & Legally blind since childhood, with more significant impairments in central vision than peripheral & 4 and 6 & Mother & Yes & Yes, over Zoom with 6-year old daughter\\

10 & 45-54 & F & Totally Blind/No Usable Vision & Some light and shadow perception, but not any usable vision; vision loss was gradual, over time & 17 & Mother & Yes & No\\

11 & 35-44 & F & Legally Blind & Vision loss is result of a stroke in 2021, now only able to see colors with pin point vision & 2 and 8 & Mother & No & No\\

12 & 35-44 & F & Legally Blind & 5-10 degrees of vision, no night vision, and cataracts, onset at age 29 & 4 & Mother & Yes & No\\

13 & 35-44 & F & Legally Blind & Stargardt’s, onset in 2014 & 7 & Mother & No & No\\

14 & 55-64 & F & Totally Blind/No Usable Vision & Some light and shadow perception & 7 and 9 & Grandmother & Yes & Yes, in-person with 7 and 9 year old granddaughters\\
\bottomrule
\end{tabular}
}
\vspace{0.5em}
\caption{Self-reported demographics of BLV study participants, including information about relationship to child and which studies they took part in.}
\vspace{-1em}
\end{table*}

\subsection{Procedure}
Study 1 consisted of three parts: (1) a pre-study questionnaire, filled out prior to the session starting, (2) a semi-structured interview, and (3) a demonstration and discussion of AI-based descriptions of children's artwork. Sessions lasted for ~60 minutes. All participants were compensated \$25.

The pre-study questionnaire collected demographics as well as information about vision loss and their children's art. For the semi-structured interview, we asked questions about how participants currently engage with their children's artwork, their motivations and desires to learn about the art and art process, and the tools and techniques they use to understand and interpret the art. 

To examine the potential for AI as an access tool for children's artwork, we invited participants to share their child-created art pieces. At the end of the study, we inputted these images into two state-of-the-art AI tools, \textit{Be My AI} \cite{be_my_ai_2024} and \textit{ChatGPT4} (\textit{i.e.}, ChatGPT using GPT-4) \cite{explore_gpts_2024}, and shared the AI output with participants. We solicited reactions to the AI descriptions, feelings about using AI to learn about their children's art, and potential opportunities and concerns. For three participants (P2, P3, P7), this AI interpretation component was conducted as a separate call after the initial session to allow for more time to collect and send us their child's artwork. 

From the many tools available for AI-generated image description, we selected Be My AI and ChatGPT4 based on a comparative examination of six AI tools conducted in February 2024: ChatGPT, BARD\cite{gemini_2024},\footnote{Google Gemini had not released at the start of this study} LLaVA \cite{llava}, Seeing AI \cite{seeing_ai}, Bing Chat \cite{bing_ai}, and Be My AI \cite{be_my_ai_2024}. For our comparison, we examined artwork from the children of two research team members (see Appendix \ref{appendix}). We found that Be My AI and ChatGPT4 most frequently provided descriptions that aligned with the parents' own descriptions. In contrast, some tools provided either erroneous or overly simplistic descriptions. For example, while Seeing AI is a well-respected BLV tool, it does not appear to have been trained for artistic descriptions and so would simply say: \textit{``This is a children's drawing of <object>,''}. See Appendix \ref{appendix} for more details about the tools' different outputs; since each tool was tuned for a distinct domain \cite{specializing_llms}, descriptions varied despite several tools being based on GPT.

\subsection{Analysis}
All interviews were audio recorded and transcribed. For analysis, we used deductive and inductive coding \cite{Braune_And_Clarke}, drawing on our interview protocol for deductive codes and affinity diagramming \cite{braun2012thematic, Jokela} of transcripts for inductive codes, thereby arriving at our initial codebook. The first author then engaged in peer debriefing \cite{PeerDebriefing} with another researcher (who was given two transcripts and the codebook) to iterate on the codes. Both researchers met to resolve disagreements, updated the codebook, and the first author then re-analyzed all transcripts with the final codes.

\section{Study 1 Findings}
We report on how blind or low-vision (BLV) family members engage with their children's artwork, their motivations and concerns, and strategies for understanding and interpretation---from conversations with the child to the use of technology and AI. Throughout, we highlight the emotional and intimate connection between the child and adult that is often cultivated through art. Quotes have been lightly edited for grammar, concision, and anonymity.

\subsection{Current Practices}
We identified five common approaches for how BLV family members engage with their children's artwork, including: most commonly, talking with the child about the artwork (\textit{N=}14/14) or asking a sighted adult partner (\textit{N=}10); other sensory approaches, such as having the child guide their hand to trace outlines (\textit{N=}4) or using tactile materials or other access adaptations (\textit{N=}4); and, finally, using technology tools (\textit{N=}3) such as \textit{Seeing AI} and \textit{Be My Eyes}. For low-vision family members who retain some sight, a common technique was to take a picture of the art with their phone and zoom in to discern more detail (\textit{N=}3).

When first presented with an art piece, all participants mentioned inviting the child to describe their artwork verbally and then asked directed follow-up questions. As P1 described: \textit{``The trick is to say, ‘Oh that's so nice. Tell me about it.’ And then they'll explain it. From that, you can ask some follow up questions like, `Which was your favorite color you used?’ By asking the questions, they tell the story.''} However, P9 and P10, both mothers, noted a balance between information seeking \textit{vs.} feeling burdensome. P10 said: \textit{``I am a blind mom, but that doesn't mean I monopolize all her time and energy to describe everything to my satisfaction.''}

Many participants asked sighted friends or family as an alternate source of information (\textit{N=}10). Some participants found it helpful to have ``adult'' vocabulary descriptions of artwork characteristics, and, accordingly, consulted with sighted adults for additional information. P4, a grandmother, recounted how her sighted husband describes their grandson’s work to communicate his artistic growth: 

\begin{quote}
    \textit{I'd like [the child’s] rendition. But a lot of times, I'll have him go show my husband. And [my husband] will tell me, ``Oh he did a really good job, [there is] a lot of balance on the paper, the buttons are in one area, and the coloring is in a certain place...'’}
\end{quote}

However, sighted friends or family do not always provide helpful descriptions for the BLV individual. For example, P3 and P10 discussed how art is subjective among adults as well, so they may not entirely capture what the child wanted to communicate. P3, an uncle, mentioned that he had experienced other adult relatives being unintentionally dismissive of children’s drawings. P2, a father, had not sought help to describe his son’s work because he did not want to seem mistrustful of his son’s descriptions: \textit{``I don't want to make him think I don't believe or understand what he's telling me. So I stick with [his descriptions] because I know he's trying his best to make sure I actually understand what he does.''}

\textbf{Beyond verbal descriptions.} Participants also employed other senses to explore artwork, such as tactility, hand guiding, and even smell. P10's daughter outlined artwork in nail polish to create a raised line so she could \textit{``feel the art.''} P7, a father, said his children often use tactile methods when creating art specifically for him: \textit{``they'll do something in modeling clay, puff paint, Wikki Stix, string art, or something along those lines where I can actually feel it.''}. P2, P4, P6, and P9 all mentioned hand guiding---their children would hold their hands to trace artwork outlines or guide their hands to a particular spot to co-create with them. Some approaches even went beyond tactility: P4 had her grandson use scented markers so she could smell the different colors they used in a drawing. Participants described their children organically adapting their artwork as they became aware of their BLV family member's disability. For example, P8, a great-grandmother, said her great-granddaughters remind each other to adapt their work: \textit{``They're really good about saying, `Oh, Nana can't see that. Get a black [felt pen] or do it in black ink.' Or [they say] `that's too small. Make it bigger. Nana can't see it.' ''}

\textbf{Use of interactive technology.} When asked about their use of digital technology, a small subset (\textit{N=}3) responded affirmatively. P11, a mother, tried \textit{Seeing AI}; P3 connected with human volunteers through \textit{Be My Eyes}; and P10 tried the AI feature of Be My Eyes called \textit{Be My AI}. Both P10 and P11 described being underwhelmed by their AI experiences. P11 said the feedback she received \textit{``This is a child's drawing''} was too simplistic and did not provide enough detail. P3 appreciated Be My Eyes because it allowed him to access an \textit{``adult description''} of the artwork, such as if the drawing was done in portrait or landscape, point of view of the work, objects in the foreground \textit{vs.} background, and more.

For those participants who had not used any technology to access their children’s artwork (\textit{N=}11), common reasons included: they had not thought to try it in this context (\textit{N=}7), they did not use accessibility apps in their day-to-day lives (\textit{N=}2), or it was easier to get sighted help (\textit{N=}2). P4 mentioned one additional technology, the \textit{Echo Show}, a home assistant with a display screen, which she used with her grandson to create artwork: 

\begin{quote}
\textit{He'll say, ``Alexa, make a picture of an elephant.’' Then he'll say to me, ``Do you want to draw a picture of an elephant?’' That way, we both are doing something together. And then he wants to write the letters up for it, [so he will] ask Alexa how to spell it.}
\end{quote}

\subsection{Motivation for Engaging with Artwork}
We also examined underlying motivations, concerns, and fears related to how BLV family members connect to and engage with children's artwork. Overall, participants emphasized the emotional and bonding value of exploring and understanding their child's artwork (\textit{N=}14/14), how it enabled them to track developmental progress (\textit{N=}4), and, for one family, how creative expression through art acted as a source of individual and collective healing. Concerning bonding, P6, a father, said: 

\begin{quote}
\textit{I think the connection there is invaluable. He created something... he could show it to anyone, and I'm on the list of people that he wants to show it to. If I fall flat and [I say] I don't know what this is... he [will think], well, that was boring. I'm not going to show dad any more artwork.}
\end{quote}

P3 echoed this sentiment of wanting children to feel satisfied with his response: \textit{``If you appear standoff-ish, chances are they will not want to come back to you again with a new piece of artwork because they were not satisfied with your response in the first place.''}

P4 and P10 both highlighted that artwork was a source of immense pride for their children, which is why it felt important for them to understand their art. P7 similarly described wanting to grow with his children’s interests \textit{``and show that I'm genuinely interested in what [they] like''}. P2 discussed that, as a single father, he cultivates a family dynamic where he can \textit{``engage in everything''}. 

Another motivation, expressed by parents more than other relatives, was to use an understanding of artwork to evaluate their children’s developmental progress. P12 has a son with Down syndrome, and she explained that: 

\begin{quote}
\textit{It would be nice to know his growth and his progress and his abilities. He has Down syndrome, and so I know his gross and fine motor skills are a little bit delayed and [understanding his art] would help manage my expectations on that side of things.}
\end{quote}

Similarly, P7 discussed the value in having an objective understanding of his children’s artwork because, as their father, he wanted to understand their development and ensure that his children were submitting their best work for school and other assignments: \textit{``Something that's a [series of] scribbles, I'm not going to submit it in the artwork that's going to be displayed in the hallway at the school.''}

P13, a mother, highlighted another motivation---engaging with her child’s art as a source of individual and collective healing. P13 opened up about her daughter being bullied at school and how art was not only a creative outlet for her daughter, but also a way that P13 could sense that her child was healing: 

\begin{quote}
\textit{I want her to be able to express and feel comfortable. I want her not to have a sense of anxiety around creating. When school first started, she would just create and she was so full of joy, but when she kind of started being in the company of other children, her perception of self changed. We decided... to [home-school] and it took almost a year, but she has resumed her joy for creating and her personality has returned. [Art] was for leisure, it was for play initially, but there was a season where, frankly, art was a part of the healing process for her... Even when she can't put things into words, she can create something really beautiful and meaningful, and it brings her comfort and joy... It's been a really healing part of our dynamic as a family.}
\end{quote}

\subsection{Preferences for Accessing Artwork} \label{study1-preferences} 

We asked \textit{what} information BLV family members desire about their children's artwork and \textit{how} they prefer to gain that information---and identified two interdependent response themes: (1) \textit{descriptive information} such as color, medium, and object location, and (2) \textit{interpretive information}, such as why the child created the art, their artistic choices, and what pieces of the art \textit{mean} to the child. The distinction between gaining descriptive \textit{vs.} interpretative information also corresponded to access preferences.

\textbf{Access method preferences.} All 14 participants valued children's verbal descriptions over any other approach because they centered the children's perspectives of their artwork, and allowed them to mix both description (\textit{e.g.}, colors used) and interpretation (\textit{e.g.}, why they used such colors and what they mean). Three of 14 adult participants emphasized that they do not need to know artwork traits beyond what their child wants to describe---and that there is so much more to art than the characteristics of the piece itself. As P12 expressed: \textit{``Just listening to the enjoyment that he gets from sharing his artwork, even though it's not super descriptive... that's enough for me.''} Similarly, P11 emphasized understanding her child's intention behind his work as more important than knowing the exact visual output:

\begin{quote}
    \textit{I feel like I don't need to understand his work as much as I need to understand why he's drawing what he's drawing. And I have no idea what his drawing is really looking like these days. But he's 8, so that's fine. He can be as good or as bad as he wants... I think that's the main point. I don't need to be able to see it, my focus in doing art with him is not to see it [myself].}
\end{quote}

P9 mentioned that she had found herself more curious about descriptive details as her daughter aged, but she prioritized intent and the importance of listening to the child's perspective: \textit{``The most important thing to me is hearing them talk about their intent and... about what they've created... [for me] the biggest piece is, `Tell me what you want me to know about your artwork.' ''}

In addition to verbal descriptions, participants emphasized the benefits of tactility, which allowed for a different experience of the artwork. For example, P6 felt tactile modalities allowed more objective details of his son’s work: \textit{``I think as he gets older, I would prefer tactile more... The representation gets more intricate and also gets more accurate.''} P4 enjoys both auditory descriptions and tactile representations, but that when doing art with her grandson (age 4), her preferred medium is tactile: \textit{``When I'm with my grandchild, I think the tactile part helps me a lot to see where he's going with his imagination.''} She also emphasized that this is situation-dependent: \textit{``When I go to the art museum, [descriptions are] fabulous, because the description is almost like painting a picture in my mind.''}

There were some concerns about tactile modes, including: worries about damaging the art during touch-based exploration if it exists in a tactile representation (P7), uncertainty about how to even start trying to represent child-created artwork in a tactile manner (P5), and the flimsy nature of DIY tactile art (P12).

\textbf{The interpretation gap.} The different access method preferences highlighted underlying tensions between description and interpretation. P3 and P7 discussed the gap between what a child describes to you and what is on the paper. P3 called this \textit{``the interpretation gap''}: 

\begin{quote}
    \textit{So the child would say, ``there's a little girl in a swimsuit,'' or ``there are little boys in swim trunks.'' And your mind is picturing an actual person because you can't see the picture. But, in effect, what they're describing are stick figures. It's the interpretation gap.}
\end{quote}

Indeed, some participants contemplated the dichotomy between hearing the child’s interpretation and a more \textit{objective} description---a tension that also emerged with using AI tools (see Section \ref{study1-AI-analysis}). As P6 explained: 

\begin{quote}
    \textit{I feel like if my [adult] friend described a piece of his art to me, it's more how I would perceive the art if I were looking at it. Because I can't parse the image, I would like to hear how someone else looks at it, because it's closer to how I would see it. And then I would like to hear how [my son] sees it because that's how he drew it, and that's what he's really excited about.}
\end{quote}

For P3 and P7, bridging the interpretation gap required access to both the child's perspective and a more objective description. P7 said that he often asks his sighted wife for the latter but can also converse with his children (ages 7 and 9) as they get older: 

\begin{quote}
    \textit{I can say, ``That's a really good idea,'' or ``Oh, are you sure, you know, you want to color that orange; shouldn't you be coloring the water blue?'' As they've gotten older and more mature, we've been able to take those conversations to... a deeper level.}
\end{quote}

P14, a grandmother, was also interested in knowing more descriptive or objective qualities, but wanted to honor the child's own narrative about the art:

\begin{quote}
    \textit{I will go with their descriptions. Because with the littler ones, when they're giving you that triangle dog, [to them] that's our family dog. Sighted people might say, but that's [just] a triangle. Well, that tears apart what the kid has got in his head... That's not very fair.}
\end{quote}

\textbf{What information to reveal.} Finally, when asked about \textit{what} objective information about a child's artwork was most important to reveal, participants stated characteristics such as color, technique of the work as well as details about intricacy of a piece.

P8 cannot see certain colors now due to her vision loss, but her great-grandchildren know that she loves colors, so she mentioned always asking them about colors on the page. Similarly, P4, who is an artist, discussed colors being a prominent part of discussions with her grandson about his artwork. Other parents, such as P10, preferred information about technique over color:

\begin{quote}
    \textit{I'm used to not dealing with color so much. But I like to know how detailed it is, or [about] what techniques [were used], like brush strokes or whatever she used to just get an appreciation for the talent and how much [effort] she put into it.}
\end{quote}

P5, a low-vision mother, wanted to know information about the intricacies of her son's artwork. Her child enjoys creating small but extremely detailed drawings, and she described: 

\begin{quote}
    \textit{When he was younger it was a little easier [to understand them], because he tended to draw larger pictures that didn't have that fine detail. Now, he’ll draw pictures that are sometimes a couple inches across, and it's a full body sketch... To have the detail of it described would probably help me to be able to see it.}
\end{quote}

\subsection{Reactions to AI Tools} \label{study1-AI-analysis}
At the end of the study, we invited participants to share their children's artwork and analyzed them with two AI tools: \textit{Be My AI} and \textit{ChatGPT4}. Ten of our 14 participants had artwork available and took part in this portion of the study (Table 1). In general, most participants (\textit{N=}9) reacted positively but also discussed the lack of emotional depth, accuracy (\textit{e.g.}, discrepancies between the two AI descriptions), and how this technology could be used in the future.

\textbf{Potential benefits.} Our participants identified several key benefits of using AI to analyze children artwork, including: the ability to discern details that were otherwise inaccessible (\textit{e.g.}, colors, painting techniques), access to information that could further conversation and connection with their child, and using AI to complement the child's description, particularly when they were not available (\textit{e.g.}, P10's former partner sends pictures of her daughter's work when her daughter is staying with him). As one poignant example, P7, a father, had taken his family on a trip to his alma mater university's football game. Some time later, his son made a drawing of the experience and shared it with his father; however, the father did not realize just how much was captured by the picture (Appendix, Figure \ref{FootballGame}), including details about the players and fans in the stands, until after hearing the AI description. P7 emotionally shared: \textit{``Hearing that [detailed AI description], I could have had a 10 minute conversation [with my son] and... appreciation for what he had done. It’s dad guilt... I [originally] dismissed [the artwork]...'''}. Here, P7 felt that having access to an AI description could have helped him better understand his child's artwork and his child's attempt to connect with him.

P2 compared the AI description with his current practice of hand-guided tracing of his son's drawings: \textit{``I get more information from it. Because [hand] tracing, it's fine. But with the AI, you actually give me whether the person [in the drawing] has something in their hand... How they dress, the facial features...''} P2 liked the prospect of not needing to wait for his son's descriptions, and instead he could use time with his son to ask specific questions about the work after it has been described by the AI. 

P14 appreciated hearing descriptive qualities of the artwork, such as the colors and whether coloring was done \textit{``inside the lines.''} Since she has many grandchildren, she was also excited about the possibility of using AI to identify who made the artwork.

\textbf{Interpretation \textit{vs.} description.} Similar to Section \ref{study1-preferences}, tensions arose around AI providing \textit{descriptions} about the artwork (\textit{e.g.}, color, medium, line thickness) and providing \textit{interpretation}. Some participants (\textit{N=}3) appreciated interpretation---such as inferring emotions on a drawn character's face (P2, P3)---others felt that the AI should avoid making inferences. For example, P9 disliked when the AI described one of the pictures drawn by her daughter as \textit{``simple''}. P12 described her irritation when the AI interpreted a green brushstroke in an abstract art piece as a tree or a plant (Figure 1c), saying: 

\begin{quote}
    \textit{I don't want a robot to tell me like, ``this looks like a preschooler trying to draw a bird'' or whatever... because I know that if I ask my son, he's not going to say, ``Oh, I made you a blue painting.'' He's going to tell me it's [his friend] eating a lollipop... He's going to have a story behind it.}
\end{quote}

\begin{figure*}[h]
\centering
\includegraphics[width=\linewidth]{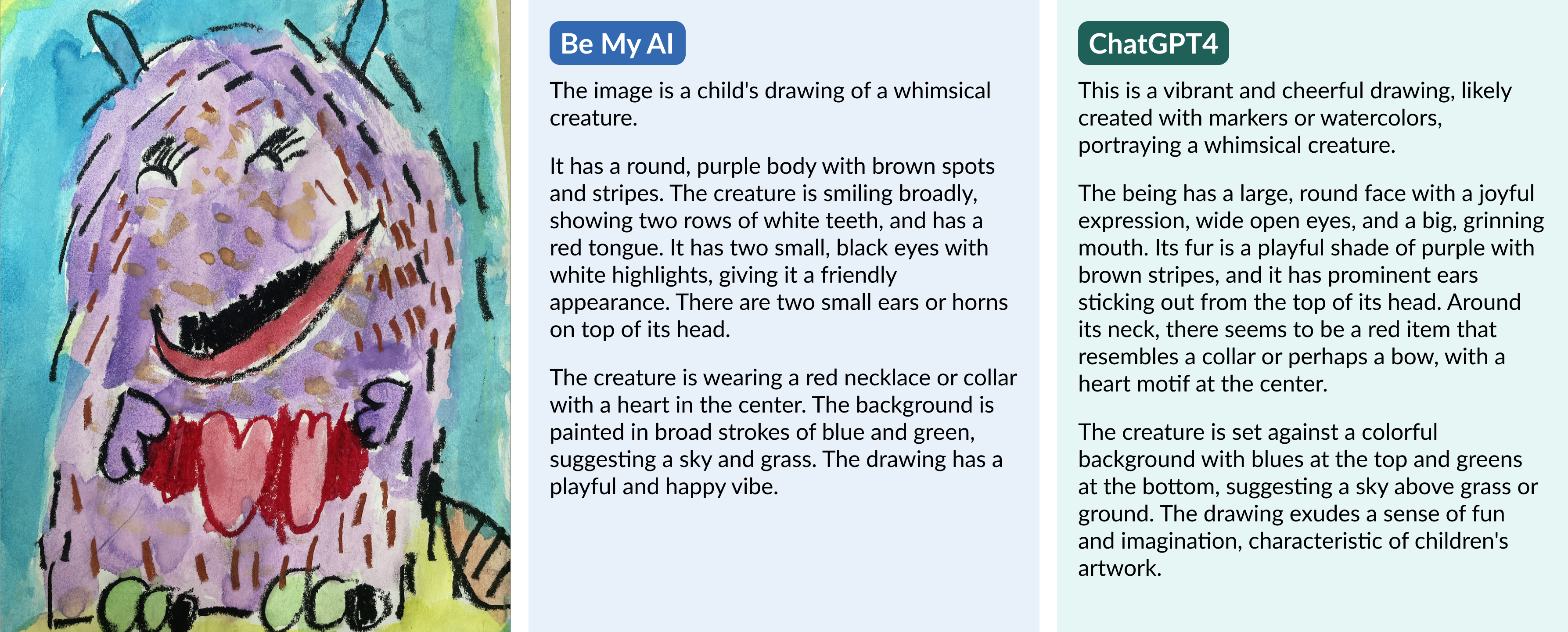}
\Description{This image is split into three panels. On the left is the artwork by the child, which has a blue background and a large purple creature covering most of the page. The creature is smiling with its eyes closed, and is holding a row of red and pink hearts. The creature has a brown tail like a beaver and green feet. The middle panel has the Be My AI description, which says: "The image is a child's drawing of a whimsical creature. It has a round, purple body with brown spots and stripes. The creature is smiling broadly, showing two rows of white teeth, and has a red tongue. It has two small, black eyes with white highlights, giving it a friendly appearance. There are two small ears or horns on top of its head. The creature is wearing a red necklace or collar with a heart in the center. The background is painted in broad strokes of blue and green, suggesting a sky and grass. The drawing has a playful and happy vibe." The right panel has the ChatGPT4 description, which says: "This is a vibrant and cheerful drawing, likely created with markers or watercolors, portraying a whimsical creature. The being has a large, round face with a joyful expression, wide open eyes, and a big, grinning mouth. Its fur is a playful shade of purple with brown stripes, and it has prominent ears sticking out from the top of its head. Around its neck, there seems to be a red item that resembles a collar or perhaps a bow, with a heart motif at the center. The creature is set against a colorful background with blues at the top and greens at the bottom, suggesting a sky above grass or ground. The drawing exudes a sense of fun and imagination, characteristic of children's artwork."}
\caption{P8's great-granddaughter's drawing of \textit{Love Monster} with AI-generated descriptions. Both Be My AI and ChatGPT4 begin with a declaration \textit{``a child's drawing of a creature''} but also intermix varying levels of interpretation (\textit{``whimsical''}, \textit{``cheerful''}, \textit{``resembles a collar or perhaps a bow''}). Though some descriptions are accurate---its color, smile, and the background--others, such as the wide open eyes, or the hearts being a necklace, are incorrect.}
\label{LoveMonster}
\end{figure*}

Participants also felt the AI descriptions did not always capture intended emotions associated with some of the artwork. For example, when hearing the AI describe her great-granddaughter's picture (Figure \ref{LoveMonster}), P8 said: \textit{``I thought [the AI] missed the joy that she put into it. And I thought it missed the love.''} 

\begin{figure*}[h]
\centering
\includegraphics[width=\linewidth]{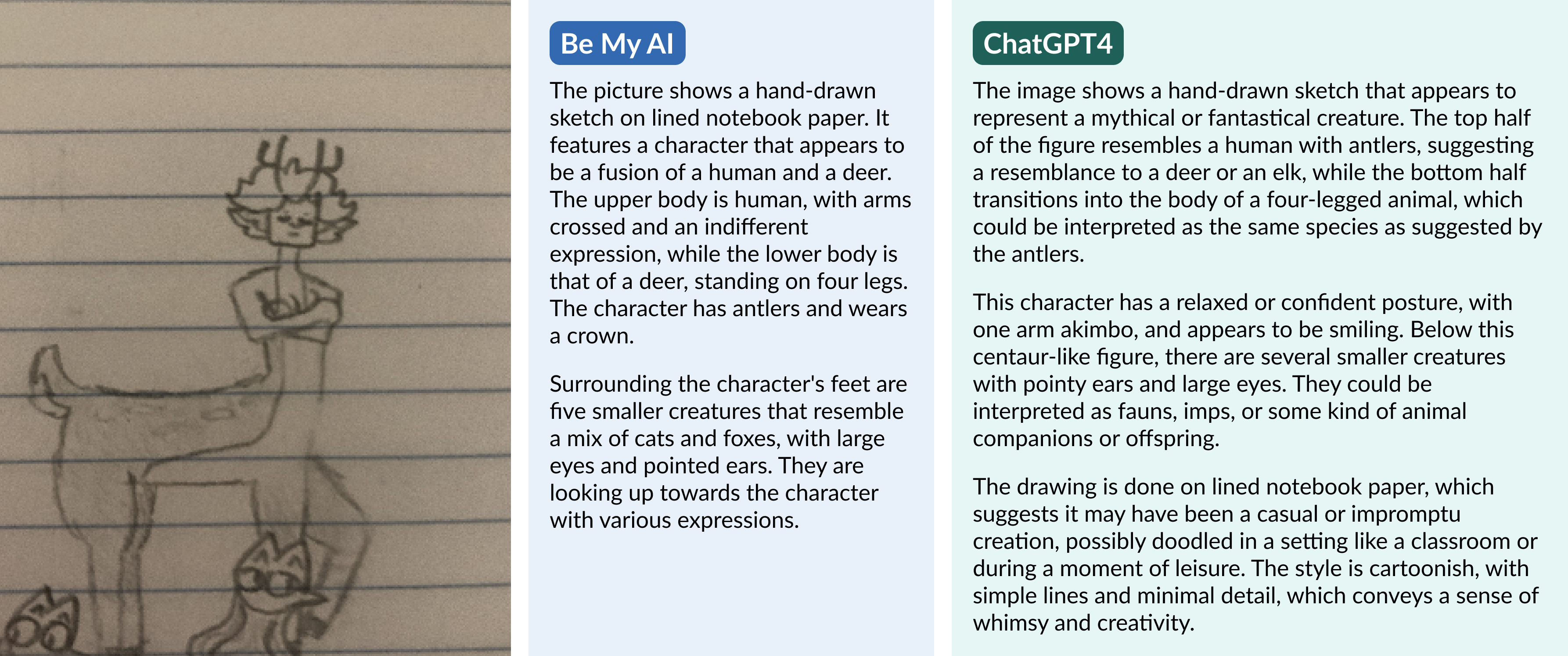}
\Description{This image is split into three panels. On the left is the sketch by the child, which is done on lined notebook paper in pencil, and has a centaur-like figure with its arms crossed and its eyes closed. It has a tail and some antlers. Below the centaur are two creatures with large eyes, and they are looking in opposite directions. One is at the bottom left of the page, and we can only see its eyes and ears. The other is towards the bottom center of the page. The middle panel has the Be My AI description, which says: "The picture shows a hand-drawn sketch on lined notebook paper. It features a character that appears to be a fusion of a human and a deer. The upper body is human, with arms crossed and an indifferent expression, while the lower body is that of a deer, standing on four legs. The character has antlers and wears a crown. Surrounding the character's feet are five smaller creatures that resemble a mix of cats and foxes, with large eyes and pointed ears. They are looking up towards the character with various expressions." The right panel has the ChatGPT4 description, which says: "The image shows a hand-drawn sketch that appears to represent a mythical or fantastical creature. The top half of the figure resembles a human with antlers, suggesting a resemblance to a deer or an elk, while the bottom half transitions into the body of a four-legged animal, which could be interpreted as the same species as suggested by the antlers. This character has a relaxed or confident posture, with one arm akimbo, and appears to be smiling. Below this centaur-like figure, there are several smaller creatures with pointy ears and large eyes. They could be interpreted as fauns, imps, or some kind of animal companions or offspring. The drawing is done on lined notebook paper, which suggests it may have been a casual or impromptu creation, possibly doodled in a setting like a classroom or during a moment of leisure. The style is cartoonish, with simple lines and minimal detail, which conveys a sense of whimsy and creativity."}
\caption{P5's son's sketch of a Centaur-like figure with two creatures at the bottom with the AI-generated descriptions. Again, both AI engines intermix descriptive and interpretative elements. They begin with the type of drawing: \textit{``hand-drawn sketch''} before describing the creature, its posture, and even its expression. For the surrounding animals, the AI-generated descriptions use suggestive language: \textit{``resemble a mix of cats''} and \textit{``interpreted as fauns, imps, ...''}. ChatGPT4 outputs more presumptive interpretation and context: \textit{``This character has a relaxed or confident posture''} and \textit{``casual or impromptu creation, possibly doodled in a classroom''}.}
\label{Centaur}
\end{figure*}

Similarly, P5 reacted positively to descriptive elements of the AI analysis but not to the AI's interpretation (Figure \ref{Centaur}):

\begin{quote}
    \textit{I understand this description, but I had a hard time... It's too sterile. Because I feel like there's more expression in the drawing... I think it was a good description of the Centaur-type character with little antlers. But it felt like it kind of lost its soul.}
\end{quote}

\textbf{Perceptions of accuracy.} Many participants seemed to understand the limits and imperfections of AI, and the use of two AI tools concurrently allowed for comparisons and more informed conversation about AI differences. Key concerns about AI accuracy included the potential for asking children about incorrect or nonexistent elements of their work (P12), fear of missing an important detail about the work (P10), and misunderstanding their child's work altogether (P5). However, some, like P7, seemed unfazed. We specifically noted that the AI miscounted the number of football players in his son's drawing, and the AI misread his son's name:

\begin{quote}
    \textit{[A mistake in the AI description] doesn't bother me one bit... It captured 95\% of it accurately. That is 94\% more than I would have gathered from what I had going into it. All I knew is this was a picture of [my son’s] interpretation when we had just gotten home from the football game.}
\end{quote}

\begin{figure*}[h]
\centering
\includegraphics[width=\linewidth]{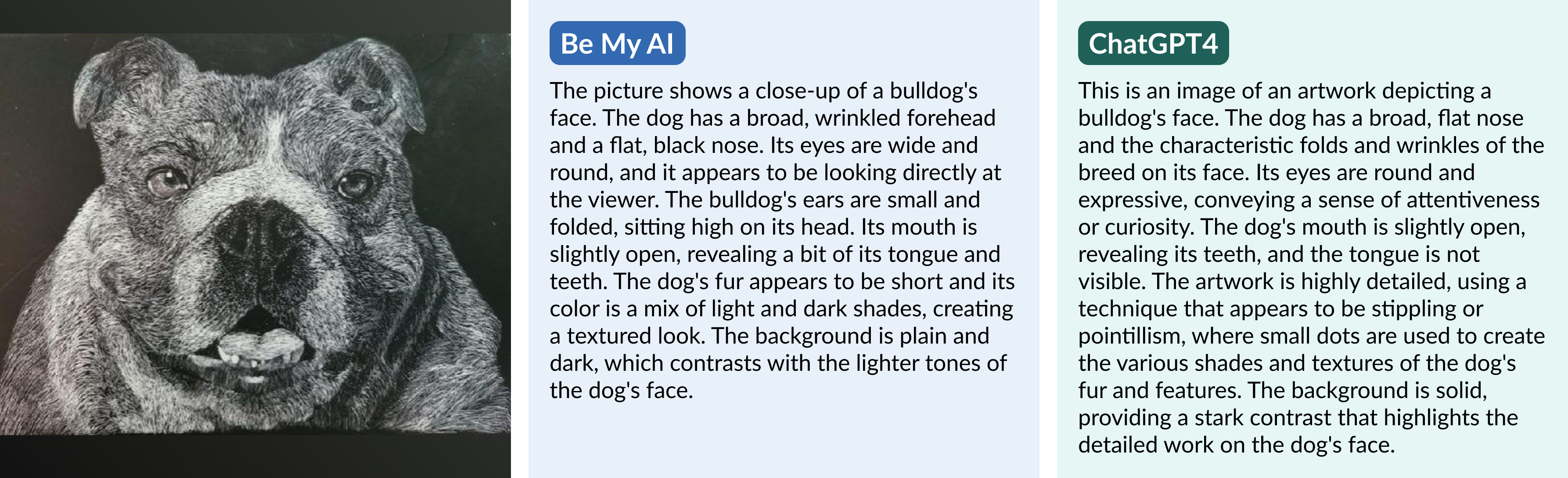}
\Description{The image has three panels, on the left is the child's artwork of the bulldog. The art looks extremely realistic and is in black and white. The bulldog is looking towards the viewer with an open mouth and raised ears, seeming happy. The middle panel has the Be My AI description, which says: "The picture shows a close-up of a bulldog's face. The dog has a broad, wrinkled forehead and a flat, black nose. Its eyes are wide and round, and it appears to be looking directly at the viewer. The bulldog's ears are small and folded, sitting high on its head. Its mouth is slightly open, revealing a bit of its tongue and teeth. The dog's fur appears to be short and its color is a mix of light and dark shades, creating a textured look. The background is plain and dark, which contrasts with the lighter tones of the dog's face." The right panel has the ChatGPT4 description which says: "This is an image of an artwork depicting a bulldog's face. The dog has a broad, flat nose and the characteristic folds and wrinkles of the breed on its face. Its eyes are round and expressive, conveying a sense of attentiveness or curiosity. The dog's mouth is slightly open, revealing its teeth, and the tongue is not visible. The artwork is highly detailed, using a technique that appears to be stippling or pointillism, where small dots are used to create the various shades and textures of the dog's fur and features. The background is solid, providing a stark contrast that highlights the detailed work on the dog's face."}
\caption{P10's daughter's bulldog art with AI-generated descriptions. Both descriptions begin with a declaration \textit{``a bulldog's face''}, then diverge in descriptive \textit{vs.} interpretive language. Be My AI describes the bulldog's eyes as \textit{``wide and round... looking directly at the viewer''}, whereas ChatGPT4 makes inferences about the bulldog's state of being from its eyes: \textit{``round and expressive, conveying a sense of attentiveness and curiosity.''} ChatGPT4 also infers the technique used as: \textit{``stippling or pointillism.''}}
\vspace{-0.5em}
\label{Bulldog}
\end{figure*}

All participants appreciated access to both AI descriptions, even if there were discrepancies. P10 described how she reconciled conflicting information from the two AI tools in relation to her daughter's artwork (Figure \ref{Bulldog}): 

\begin{quote}
    \textit{I think they're both useful to me... The second one mentions the attentive expression of the eyes, even though the other one said it appears to be looking at you, which basically conveys the same. But the first one said teeth and tongue [were visible], and the second one said the tongue isn't visible, so now I don't know if the tongue is visible or not... But at least I know the mouth is open.}
\end{quote}

Others emphasized that the two AI tools seemed complementary, providing different pieces of information. P6 said: 

\begin{quote}
    \textit{Even if one of them was more interpretive, it would describe elements that the other would miss. So I really appreciated hearing them both... I wish I had a simple platform where I could just upload them [the drawings] to both [AI tools] simultaneously, and it would just give me [both] descriptions without having to go separately and import [the art] into each app.}
\end{quote}

\textbf{When to use AI.} After sharing the AI analyses, we asked participants about whether and how they might envision using AI tools with their children's artwork in the future. Nine of 10 participants thought that they would, either before interacting with the child so they could ask specific questions (\textit{N=}4) or afterwards to improve their own understanding and recall of the artwork's characteristics (\textit{N=}5). P6 also discussed the possibility of using an AI description tool in the middle of the interaction with his son. He said: 

\begin{quote}
    \textit{[I could imagine] making a game out of it. First having the AI describe it, and then playing [with my son] and saying, ``Well, I think it looks like a cat''... and then we could [playfully discuss] whether it's a cat or a turkey.}
\end{quote}

\section{Study 2: Children and BLV Family}
To complement Study 1, we conducted a follow-up study with a subset of Study 1 participants along with their children (Table 1). Here, our goal was to focus on the child and their process of creating and sharing their art with the blind or low-vision (BLV) adult: how did the child think about their creations? Were their artistic choices, materials, or method of sharing influenced by a potential BLV audience? How did they feel about and react to AI interpretations of their work? We additionally wanted to observe the deliberate or implicit practices that occur when mixed visual-ability families connect over artwork and families' reactions to technological design probes representing artwork nonvisually.

\subsection{Participants}
Five BLV adults from Study 1 agreed to participate in Study 2 with their child: P6, P7, P8, P9, and P14. For each session, the number of children with the adult ranged from one to three (Table 1). P8 and P14 participated in person with their families while P6, P7, P9 and their respective children participated virtually. P14 asked to do Study 1 and Study 2 as consecutive in-person sessions on the same day, so we incorporated a break to avoid fatigue.

\subsection{Procedure}
Sessions consisted of either two or three parts: (1) a semi-structured interview of the child(ren); (2) an AI-based analysis of the children's artwork followed by a discussion; and, for the in-person sessions, (3) a design probe. As with Study 1, we invited participants to share pictures of art pieces made by their children. Sessions lasted 30--90 minutes, and participants were paid \$25 per hour for remote participation or a home visit, or \$40 per hour if they traveled to us.

During the semi-structured interview, we asked each child about how they share their artwork with their BLV family member, tools they use to create artwork, and what they wanted family members to know about their artwork. We leveraged artwork that participants shared with us by asking children to describe their art to their BLV family members. 

For the AI interpretation, the lead researcher inputted the provided artwork into \textit{Be My AI} and \textit{ChatGPT4}, and shared the AI output with the family. We solicited the children's reactions to the AI descriptions, and their feelings around their BLV family member using AI tools to understand their art. 

The two in-person studies ended with a design probe. This probe consisted of tactile, touchscreen, and audio representations of two different sample children's drawings collected from the authors who have children. One tactile representation used laser cutting, and another used raised dot stickers similar to swell paper; both are effective techniques used for nonvisual tactile explorations of artwork \cite{Holloway_MakingSenseOfArt}. The touchscreen component utilized Seeing AI's ``Explore'' mode, allowing spatial exploration of the artwork \cite{seeing_ai}. Figure \ref{DesignProbes} in the Appendix shows the drawings we used along with their nonvisual representations. After the adult and child tried each of the probes, we asked follow-up questions to capture their reactions to the different media. Owing to the tactile aspect of these design probes, we did not conduct probes for our Zoom sessions.

\subsection{Analysis}
As with Study 1, all Study 2 interviews were audio-recorded and transcribed. We used similar analysis methods for both studies, including inductive and deductive coding, affinity diagramming to arrive at an initial codebook, and iterating on the codebook through peer debriefing for the final codes---the first author connected with the same researcher as for Study 1 analysis, presenting two Study 2 transcripts and the initial codebook.

\section{Study 2 Findings}
We describe how children share their artwork with blind or low-vision (BLV) family members, including a distinction between \textit{context} and \textit{content}, how children react to hearing AI-generated descriptions of their art, and family responses to our design probes.

\subsection{How Children Share Their Art}
When children share their artwork with BLV family members, we found that they describe two inter-related attributes: the \textit{context}---the inspiration for the art, its intention, where it was created, if it was co-created with another person (a sibling, a teacher), \textit{etc.}; and the \textit{content}---characteristics of the artwork itself, such as color, shapes, and objects. For both, children reveal otherwise implicit details related to the personal meaning of the art---from the underlying story of a painted scene to the inclusion of their favorite cartoon character or the BLV adult's favorite color.

\begin{figure}[h]
\centering
\includegraphics[width=\linewidth]{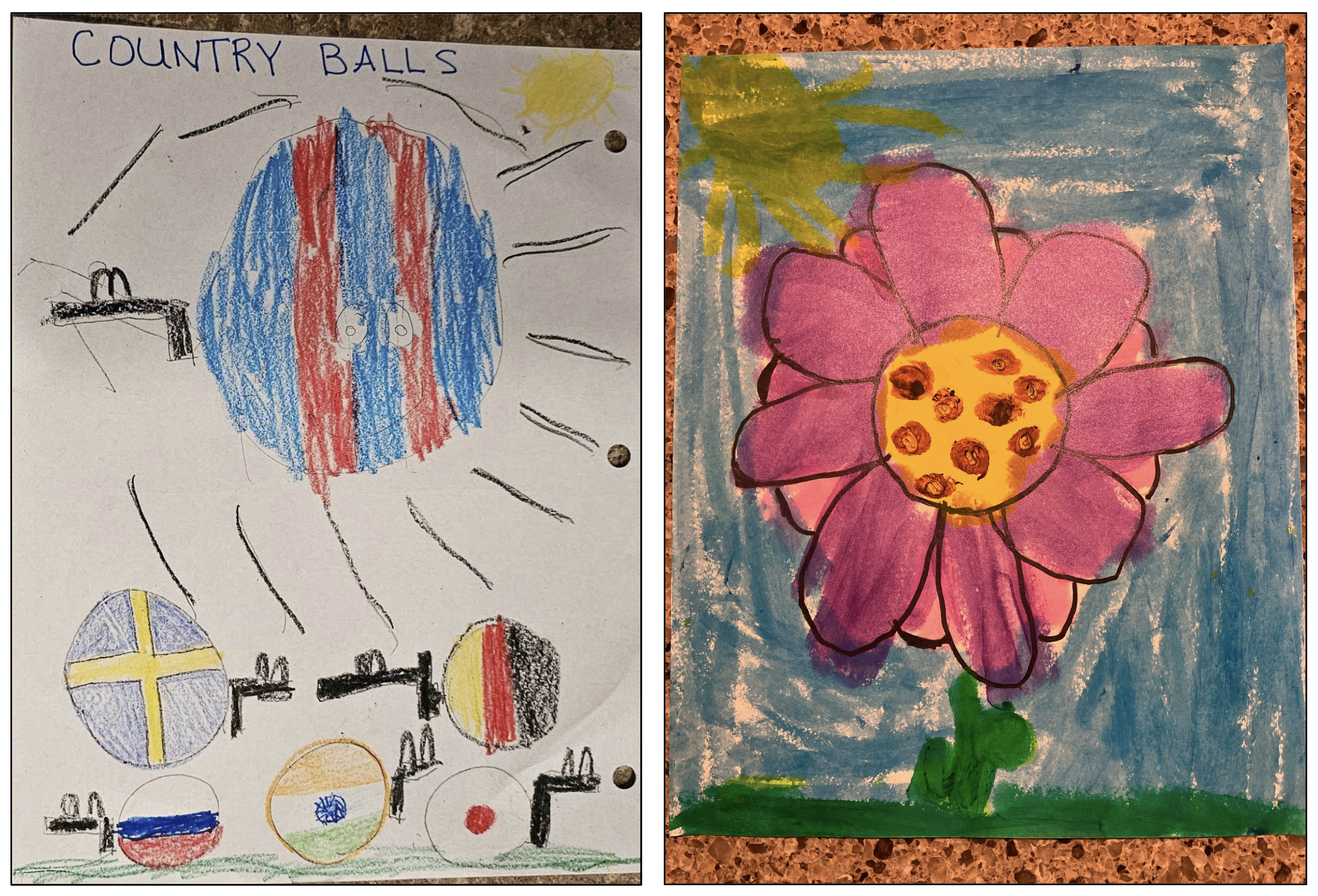}
\Description{Two images in portrait mode next to each other. On the left: a colored-pencil drawing titled "Country Balls" that has a large ball with blue and red stripes in the center of the page, and 5 other smaller balls below that large ball. The smaller balls (from left to right, top to bottom) are: a blue ball with a yellow stripe i.e. Sweden, a ball with yellow, red, and black stripes i.e. Germany, a ball with white blue and red horizontal stripes i.e. Russia, a ball with saffron, white, and green horizontal stripes with a blue circle in the middle i.e. India, and a white ball with a red dot in the center i.e. Japan. Each ball has a thin black gun attached to it. On the right: a crayon and water color image of a pink flower with a yellow center and brown spots, on a green stem. The background is blue, and in the top left corner there is a yellow sun.}
\caption{Drawings by P6's son (\textit{left}) and P7's daughter (\textit{right}). The children's descriptions emphasized both \textit{context} (inspired by a YouTube channel; made for a school art event) and \textit{content} (flag-themed balls; flower with seeds and big purple petals).}
\label{CountryBallsAndFlowers}
\vspace{-0.5em}
\end{figure}

\textbf{Sharing context.} For children, conveying the context of the art was just as important as the art itself. For example, P6’s son began his description of a drawing (Figure \ref{CountryBallsAndFlowers}), \textit{``I used to watch this YouTube channel. It had country balls in it, and they were fighting, and I got really interested in it. So I drew the poster. It's still in progress.''} Such details enabled P6, the father, to ask follow-up questions, such as which parts were still in progress. Similarly, P14's granddaughter excitedly explained to her grandmother that her artwork involved a dog from her favorite show, ``Paw Patrol.'' 

Of course, descriptions often seamlessly intermixed both content and context. For example, P7's daughter first described the content before providing context: 

\begin{quote}
    \textit{It's a flower with seeds in the middle and has purple big petals and little pink outlines of another row of petals, but behind it. I actually made that one in art class... It also was in ``Art to Remember.'' So we got magnets of that same picture.}
\end{quote}

After offering her description, P7’s daughter found and shared the magnet with us and her father, indicating the importance of that contextual detail for her. 

Context also included nonvisual identifiers of the artwork, such as P8's 7-year old great-granddaughter naming a creature she drew, ``\textit{Love Monster}'' (Figure \ref{LoveMonster}), and intention behind the art. P8's great-granddaughter said about the ``Love Monster'': \textit{``I just wanted her [my great-grandmother] to have art... I made it purple because that’s her favorite color, and I made a Love Monster because I love [her].''}

The BLV adults noted how they valued preserving the association between the intention or inspiration of the work and the content of the art itself. P8 reflected on this after hearing a description from her great-granddaughter about an artwork (Figure 1g): \textit{``She did the beads because she knows I love jewelry, and she colored the figure brown to represent her Nana; it was really cool.''}

\begin{figure*}[!b]
\centering
\includegraphics[width=\linewidth]{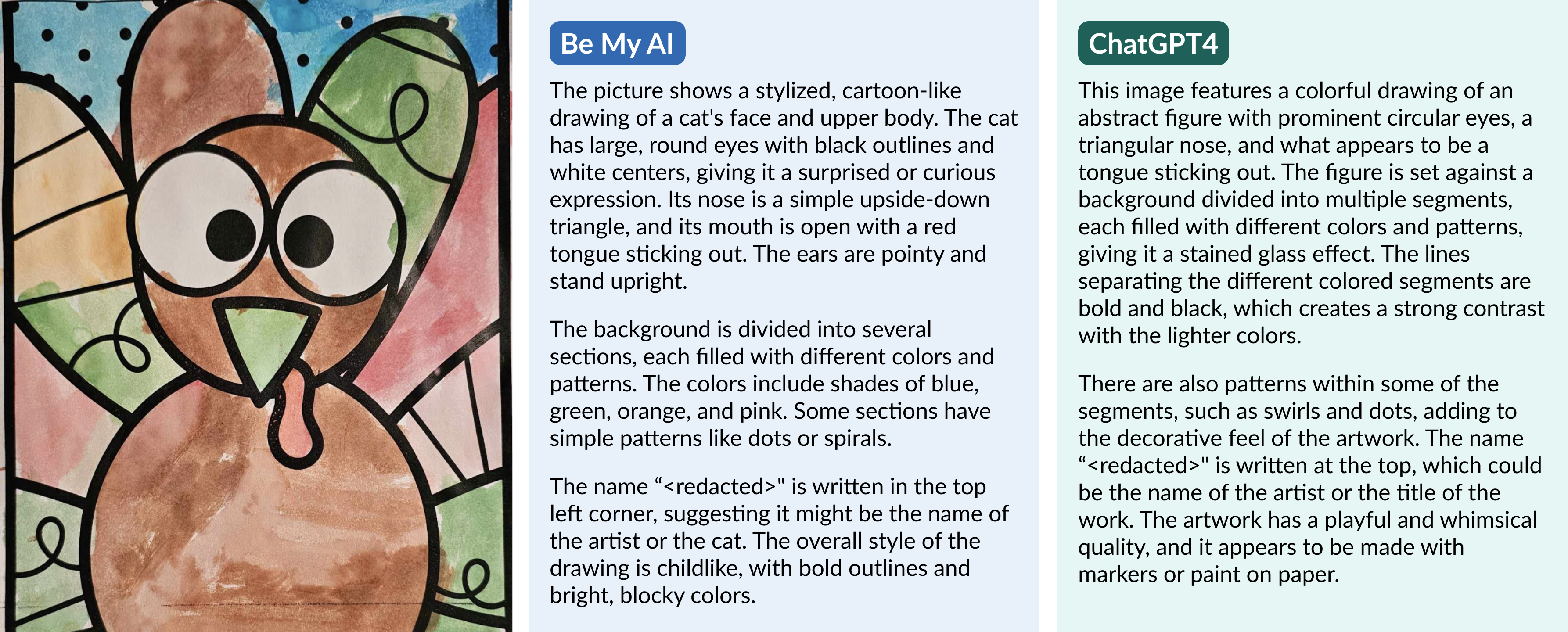}
\Description{The image has three panels, on the left is: A cartoon turkey out of a coloring book, outlined in black.  The eyes of the turkey are large and goofy. The turkey has a green beak, and a red snood. The body and face of the turkey is brown, and the turkey's tail is colored in different colors including red, green, yellow, and brown. The middle panel has the Be My AI description, which says: "The picture shows a stylized, cartoon-like drawing of a cat's face and upper body. The cat has large, round eyes with black outlines and white centers, giving it a surprised or curious expression. Its nose is a simple upside-down triangle, and its mouth is open with a red tongue sticking out. The ears are pointy and stand upright. The background is divided into several sections, each filled with different colors and patterns. The colors include shades of blue, green, orange, and pink. Some sections have simple patterns like dots or spirals. The name “<redacted>" is written in the top left corner, suggesting it might be the name of the artist or the cat. The overall style of the drawing is childlike, with bold outlines and bright, blocky colors." The panel on the right has the ChatGPT4 description which says: "This image features a colorful drawing of an abstract figure with prominent circular eyes, a triangular nose, and what appears to be a tongue sticking out. The figure is set against a background divided into multiple segments, each filled with different colors and patterns, giving it a stained glass effect. The lines separating the different colored segments are bold and black, which creates a strong contrast with the lighter colors. There are also patterns within some of the segments, such as swirls and dots, adding to the decorative feel of the artwork. The name “<redacted>" is written at the top, which could be the name of the artist or the title of the work. The artwork has a playful and whimsical quality, and it appears to be made with markers or paint on paper."}
\caption{P6's son's coloring of a cartoon turkey with two AI-generated descriptions. Both begin with very different inferences (or lack thereof) about the turkey: Be My AI calls the turkey a \textit{``cat''}, which bothered P6's son, while ChatGPT4 describes it as \textit{``an anstract figure with prominent circular eyes...''}. Both tools again intermix descriptive and interpretive traits---\textit{``surprised or curious''}, \textit{``stained glass effect''}, and \textit{``playful and whimsical''}.}
\label{TurkeyOrCat}
\end{figure*}

\textbf{Sharing content.} In the previous quote, we noticed P8's great-granddaughter explicitly identifying that the ``Love Monster'' was colored purple because that is P8's favorite color. This expression serves as an example of children sharing descriptive qualities of the art based on its personal meaningfulness and its relation to BLV family members---in this case, the color purple. As another example, P14's 9-year old granddaughter described her ``Paw Patrol'' artwork by primarily focusing on colors and clothing of the dog in the image. When asked why she did not describe some other elements, such as a border of pumpkins, she said those details did not matter to her. Her grandmother, P14, reflected: \textit{``She's focusing on what's important to her, which is the dog.''}

\subsection{Children's Reactions to AI Tools}
When hearing AI-generated output about their artwork, children preferred descriptions capturing aspects of the artwork they cared about, disliked and tried to rationalize when art elements were described incorrectly or omitted, and wanted to change inaccurate AI descriptions before their BLV relatives heard them.

\textbf{Preferences within descriptions.} What children appreciated about AI descriptions directly followed from their personal interests and artistic preferences. As an example, if one AI description expressed details about all of the colors on a page correctly, a child who was more interested in colors, such as P14’s 7-year old granddaughter, preferred that description; but a child who cared more about materials used in the artwork, such as P8’s 9-year old great-granddaughter, expressed a preference for explanations of materials and techniques (oil painting, beads) more than colors.

\textbf{Responses to inaccuracies.} When AI descriptions were mostly correct according to the children, children responded neutrally to those descriptions using vocabulary like \textit{``good''} or \textit{``correct.''} In contrast, all of the children swiftly and animatedly called out when AI interpretations got details wrong. P6’s son did so when responding to AI descriptions mistakenly calling a turkey a cat in one of his drawings (Figure \ref{TurkeyOrCat}): \textit{``[It is] totally incorrect... I hate that AI dude! Cause that is not a cat! It is not a cat.''}

In some instances, children tried to rationalize why the AI descriptions made incorrect assumptions. One of P14’s granddaughters described why she thought the AI inaccurately inferred the position of a dragonfly in relation to a cat in one of her drawings: \textit{``I can kind of tell why [the AI thought what it did]... That could be like a part of [the cat's] arm.''} As another example, both AI tools inferred it was raining in an image drawn by P9's daughter due to the presence of blue clouds. P9 and her daughter discussed the AI's inference, saying \textit{``the computer was taking a guess''} since rain is typically blue. However, P9's daughter explained that she only used blue because she \textit{``didn't have a white crayon.''}

P8’s great-grandchildren commented on the descriptions of the ``Love Monster'' (Figure \ref{LoveMonster}) incorrectly identifying some details, such as the monster’s eyes being \textit{``wide open''} when they were, in fact, closed, and missing other details altogether, such as its feet and tail. When asked how they would feel if P8 had access to the two raw AI responses of their work, the children expressed that they would want to correct the descriptions before P8 heard them. 

\subsection{Responses to In-Person Design Probes}
In our two in-person sessions with P8 and P14 and their children, we used design probes (Appendix, Figure \ref{DesignProbes}) featuring audio, tactile, and touchscreen renditions of children's artwork. Responses indicated a preference for tactile and auditory representations over touchscreens and revealed how they envisioned using each type for artwork exploration. When comparing tactile and touchscreen representations, P8 described why touchscreens are challenging: 

\begin{quote}
    \textit{I know a lot of people that are low vision have issues with neuropathy, like I do in my fingers... and so it's hard to get the right pressure over something, even with Voiceover [a mobile screenreader]. I hate [touchscreens] for that reason... If it were the tactile piece that has a little more definition in the feel of it, that would help.}
\end{quote}

P14's grandchildren helped orient her with the two tactile probes by guiding her hand over the work and describing the parts of the work as her hand reached each portion. The grandchildren described wanting to use a similar combination of tactile and verbal descriptions to share their own work with P14. P14 highlighted the importance of having the audio overview and tactile representation: \textit{``I like both because the verbal will tell you specifically what it is, but the tactile gives you a representation of what you're not seeing.''}

Between the two tactile representations---a laser cut with pieces cut out and a drawing with an outline of raised beads---P8, P14, and their respective children all preferred the raised beads. P14's granddaughters said they liked both the audio and the tactile renditions, but that they would pick tactile representation if they had to choose one medium for sharing their artwork with P14. Similarly, P8's 7-year old great-granddaughter described wanting P8 to be able to feel her ``Love Monster'' artwork (Figure \ref{LoveMonster}). However, the great-granddaughter mentioned she wanted P8 to know about the color, since purple was P8's favorite color, and she questioned how tactility would convey that information.

\section{Discussion}
Our two studies revealed current practices in mixed visual-ability families to engage with children's artwork, the information BLV family members wanted to access, as well as feedback about different technologies that might play a role in experiencing artwork. We situate these findings within prior work, discuss implications for technology designed to support these family interactions, and consider limitations and opportunities future work. 

\subsection{Revisiting Our Research Questions}
Our two studies explored three high-level questions: (1) How do BLV family members currently engage with their children's artwork? (2) What information do they desire and why? and (3) What is the role of technology in aiding comprehension and interpretation? We revisit these questions below.

Towards the first, we identified varied practices that centered the child and their verbal descriptions, which included distinctions between \textit{content} (the \textit{what}) and \textit{context} (the \textit{why} and the \textit{how}). Some families also complemented verbal descriptions with hand guiding and tactile materials---affirming the use of multi-sensory art interpretation methods in the literature \cite{Butler_SystematicReviewOfTactile, Shi_MagicTouch, Holloway_MakingSenseOfArt, Butler_MultiSensory}. Interestingly, we found that most BLV family members did not currently use technology with artwork; instead, families centered the child's perspective and storytelling. Many had not even considered using technology to explore and experience their children's art. In the few previous instances families did use systems such as \textit{Seeing AI}, they found the descriptions overly simplistic and not meaningful. 

For information preferences and rationale, we found that BLV family members considered the child's story behind the art as the most important to understand, which is why they centered children's perspectives of their artwork above other nonvisual representations. Indeed, a few BLV family members did not even care to learn about aesthetic details of the child-created work. In contrast, prior research has primarily focused on adult-created art in museums or professional settings, where BLV individuals displayed a stronger interest in understanding visual aesthetic details \cite{Li_UnderstandingVisualArtsExperiences, Asakawa_IndependentMuseumExperience}. When BLV family members did want descriptive characteristics, they were curious about the aesthetic quality of the work (colors, balance across the page, techniques) to help them understand their children's development and to appreciate the child's effort in making the art. Prior work by \citet{Li_UnderstandingVisualArtsExperiences} found that BLV individuals were motivated to learn about visual artwork for cultural learning, activism, social engagement, and relaxation. We extend these findings towards BLV family members' motivations for engaging with children's artwork, which were primarily for emotional connection and to understand children's developmental milestones.

Finally, through the AI-based analyses and design probes, we found that AI descriptions were generally considered helpful by BLV family members, and that participants (adults and children) preferred audio and tactile representations of artwork. Our results around usefulness of AI in mixed visual-ability families extends prior findings by \citet{Zhang_StoryBuddy} which highlight the benefits of AI in parent-child settings. Our results around audio and tactile being preferred by BLV individuals for accessing artwork also reaffirms findings from prior work \cite{Ahmetovic_TouchScreenExplorationOfArtwork, Li_UnderstandingVisualArtsExperiences}. In Study 2, we were able to directly observe children's reactions to the AI descriptions, which unveiled mixed reactions. Recent work by \citet{Bai2023ParticipatoryDO} describes the unique opportunity for interaction design between emerging AI agents and children, which our study further reinforces---the children in our studies were curious about AI interpretations of their work and wanted to play an active role in correcting AI descriptions when they were inaccurate. To our knowledge, our work is the first to capture children's perspectives on AI descriptions in support of their BLV family members.

\subsection{Implications for Design}
Drawing on our findings, we recommend the following design considerations for future technology to support the value of relational bonding over child-created artwork, nuances of art access, and different mediums for art representation, while reserving AI considerations for the next section.

\textit{Relational support.} Our findings build on prior work in art and child development \cite{importance_for_early_development_2024, Lynch_2012, Rymanowicz_2021, how_parents_can_help_2024}, underscoring that a child's art is not just about the artwork itself but about the child's psychological and motor development, the intimate stories and context surrounding the art, and the personal connections between child and relative. Tools that enable access to child-created artwork should not just support objective understanding and interpretation, but also the relationship between child and BLV family member.

\textit{Technology complements storytelling.} Technology cannot and will not replace a child's perspective of their artwork---their personal choices of color, of theme, of motivation. Thus, any technology should be considered complementary to the child's perspective. Still, we found a desire by BLV family members to support independent analysis of their child's artwork for personal curiosity and to further nurture conversations with their children.

\textit{The nuances of art access.} Our work identifies distinctions between \textit{content} and \textit{context} and \textit{description} vs. \textit{interpretation}. In prior work observing BLV individuals' preferences for professional or museum artwork, access to content and description (where description could include professional interpretation by art critics or the artist) mattered \cite{Li_UnderstandingVisualArtsExperiences}. Our work revealed the added nuances within mixed visual-ability family settings---content, context, description, and the child's own interpretation all hold value for how BLV family members understand their children's artwork.

\textit{The stages of access.} We observed three stages of access centered around the interaction with the child: before, during, and after. BLV family members desired access to information about the artwork both before and after their conversations with children to allow independent exploration and understanding and to better support discussion. For use during an interaction, most BLV family members did not want to distract from the moment with their child by listening to AI descriptions of their work, though P6 imagined tools like AI could be fun to use in a game-like manner. Families did see value in co-exploring tactile representations during the interaction.

\textit{The importance of tactility.} We found broad support for tactile explorations, similar to prior work \cite{Holloway_MakingSenseOfArt, Phutane_2022}; however, current easy-to-access technology solutions either provide auditory descriptions (\textit{e.g.}, voiceover descriptions from Be My Eyes \cite{be_my_eyes_2024}) or touchscreen and audio (\textit{i.e.}, Seeing AI \cite{seeing_ai}, ImageExplorer \cite{Lee_ImageExplorer}). We encourage future research into lowering the barrier for automatic tactile representations, including new interactive tactile touchscreens \cite{XuTeslaTouchCHI2011}, 3D-printed representations \cite{Stangl3DPrintIDC14}, or even manual DIY toolkits for children as created by \citet{Bae_TouchingInformation} for data literacy.

\textit{Supporting hand-guiding.} Building on the above, children often used hand-guiding in combination with verbal explanations with BLV family members. While prior work suggests multi-modal artwork explorations to enhance understanding \cite{Phutane_2022, Li_UnderstandingVisualArtsExperiences, Ahmetovic_TouchScreenExplorationOfArtwork}, we extend this to consider multi-user exploration where the child guides the BLV family member's hand and provides verbal explanation. Future technologies could either mimic this practice with input from the child or record these explorations for \textit{post hoc} use by the BLV adult.

\subsection{AI-based Design Considerations}

We extend our recommendations to consider the role of AI in supporting mixed visual-ability families, given the recent advancements and proliferation of AI tools for BLV individuals. 

\textit{Optimize for personal preference factors.} BLV individuals have personal preferences regarding interpretive and descriptive characteristics  of child-created artwork, such as color or technique. Children additionally get excited by highlighting noteworthy artwork elements for their BLV family members. AI tools should support preference inputs from the BLV adult and their children to craft tailored descriptions of artwork. For example, a description could be more or less interpretive, or focus mostly on colors used, depending on the interests of the family.

\textit{Support child corrections of inaccurate descriptions.} Children are proud of the artwork they produce and dislike when AI tools incorrectly describe their work. They also want to correct inaccurate descriptions before their BLV family members hear them. Human-in-the-loop systems for correcting AI exist, and prior work explores inclusive applications of such systems \cite{FloresSaviaga_InclusivePortraits}; we recommend making these correction workflows accessible for children to support fixing inaccurate AI descriptions for their BLV family members.

\textit{Utilize multiple models for artwork descriptions.} BLV individuals get different information from different AI tools, and they prefer accessing all of these descriptions to build a holistic understanding of their children's artwork. AI description tools should enable this with ease for BLV individuals---as an example, a single unified system could act as a front-end to multiple AI models, returning different descriptions together for simultaneous consumption, similar to a news aggregator.\footnote{For example, AllSides provides side-by-side news stories from the political, left, center, and right. See https://www.allsides.com/unbiased-balanced-news.} 

\textit{Avoid uncertain or minimizing language.} BLV family members dislike when AI reduces the significance or effort of an element of their children's artwork by using terms like ``simple'' to describe it. Similarly, AI inferences drawn about an element, such as interpreting an abstract green brushstroke by P12's son as ``could be a plant or a tree'', can detract from the BLV family member's experience with their child's work. AI tools should avoid making judgments or overreaching inferences about children's artwork. 

\subsection{Limitations and Future Work}
A limitation of our present work is that we could only conduct the in-person design probes with two BLV families, neither of which included a parent-child pair (instead, they were grandparent-grandchildren groups). We designed flexible studies \cite{Mack_AnticipateAndAdjust} to support access needs for BLV adults as well as minors, and many people preferred participating remotely. However, additional in-person design probes might uncover further insights into the preferences and motivations that BLV parents and relatives have for experiencing their children's artwork. We also noticed that younger children often felt shy in the semi-structured interview portion of our second study, which could have affected the perspectives they shared. Additionally, we lacked language and region diversity in our participant pool, since we only recruited English-speaking families within the United States. This could affect the relevancy of our findings to other cultures or geographies.

Our project and its findings open numerous avenues for future work to support mixed visual-ability families. Observational field studies of BLV relative-and-child artwork engagement would yield data in more natural, ecologically valid settings. Systems might be built that leverage AI to support aspects of the BLV adult-child experience of artwork, with features for description \textit{vs.} interpretation, correction of AI, aggregation of sources, interrogation, and and drill-down. We also encourage more investment in AI literacy for mixed visual-ability families as AI becomes more prominent. 

\section{Conclusion}
In this work, we conducted two complementary studies to explore how mixed visual-ability families engage with and experience their sighted children's artwork, capturing both the perspective and practices of the children artists as well as their blind or low-vision (BLV) family members. We found that BLV family members value the intimate and emotional practice of engaging in conversations with their children about their artwork, and prize their children's perceptions of their art above other descriptions. In terms of alternate access of child-created artwork, our two design probes showed that BLV family members and their children preferred generated audio descriptions and tactile representations. Additionally, many BLV adults stated that as their children age and their artwork becomes more intricate, tactile would be a meaningful way to consume that work. We also found that though there are reasons why BLV family members want more objective descriptions of children's artwork, such as tracking their children's developmental progress or enabling independent exploration of the art for deeper dialog with their children, any technology supplementing this family dynamic needs to prioritize the perspective and storytelling of the child about their own work. Centering the child's narrative also means that AI tools need to avoid reductive language about children's artwork, and enable workflows such as child-supported corrections for mixed visual-ability families. Overall, we believe there is great potential for AI and tactile technologies as supplementary tools in mixed visual-ability families' interactions around children's artwork.

\begin{acks}
We thank Jaden Wang and Chu Li for their help. This work was supported by the Wilma Bradley Endowed Fellowship through the Paul G. Allen School of Computer Science and Engineering at the University of Washington.
\end{acks}

\bibliographystyle{ACM-Reference-Format}
\bibliography{references.bib}

\clearpage
\onecolumn

\appendix
\section{Appendix} \label{appendix}
\subsection{Additional Artwork}
\begin{figure}[h]
\centering
\includegraphics[width=0.95\linewidth]{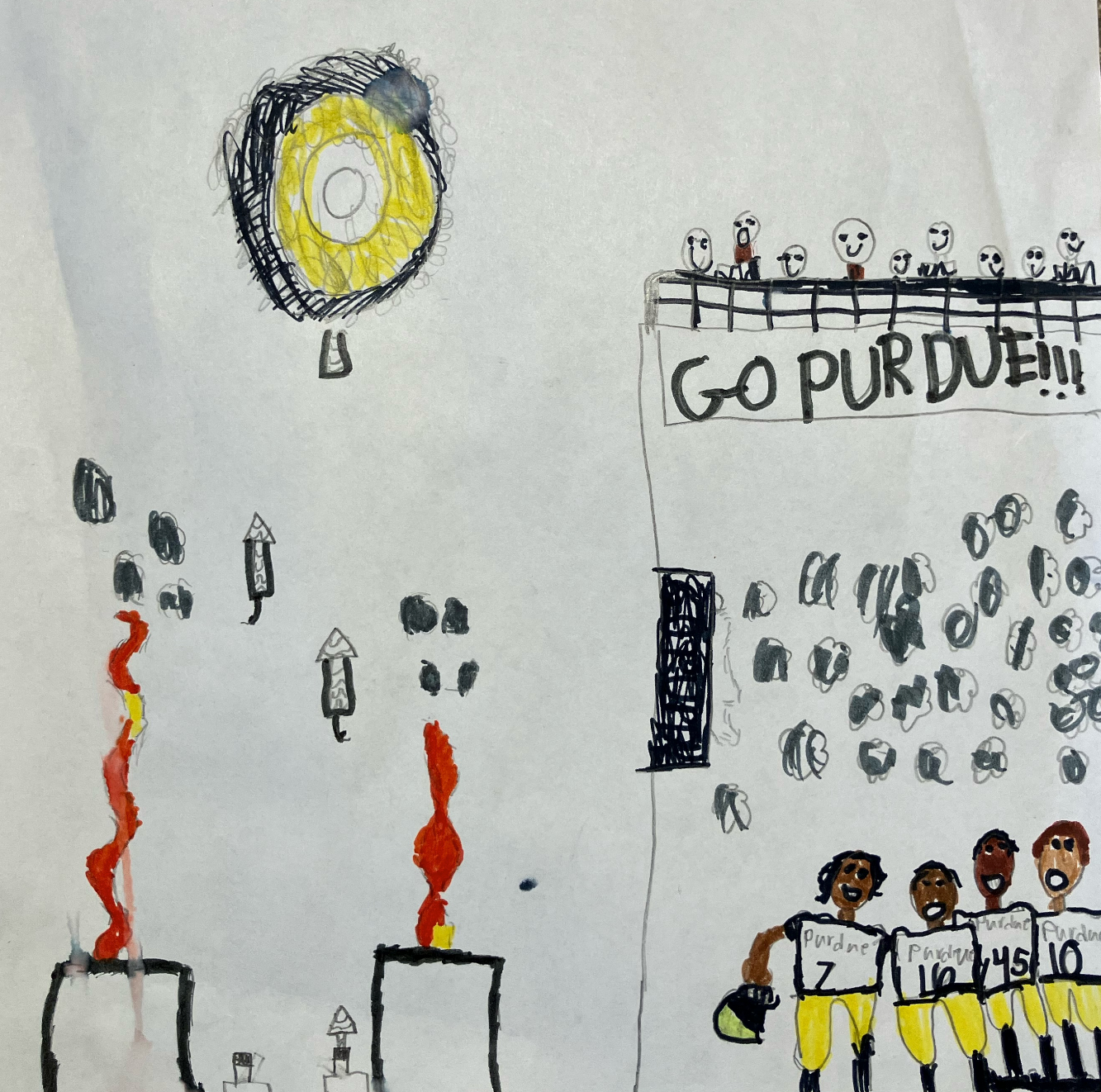}
\Description{A drawing of a Purdue University football game. At the top left of the artwork is a bright yellow light in a circular shape, outlined in black. Below that, also on the left, there are two fireworks rising to the sky, and fire and smoke coming out of two pillars. This takes up the left half of the image. In the right half of the image, there are stands near the top of the page with spectators smiling above a sign that says "Go Purdue!!!", what appears to be a football field with a goalpost section and many circles, and at the bottom of the page there are 4 players standing in close proximity to each other in football uniforms. From left to right, the numbers of the players are: 7, 16, 45, and 10. Each player is wearing yellow pants, colored in shades of brown, and the leftmost player is holding a green helmet.}
\caption{P7's son's drawing of a Purdue University football they attended as a family. The artwork captures many elements that P7 was initially unaware about, such as the opening fireworks, players running onto the field, the crowd in the stands, and the stadium spotlight.}
\label{FootballGame}
\end{figure}
\vspace{1.0em}
\clearpage
\onecolumn
\subsection{Design Probes}
\begin{figure*}[h]
\centering
\includegraphics[width=\linewidth]{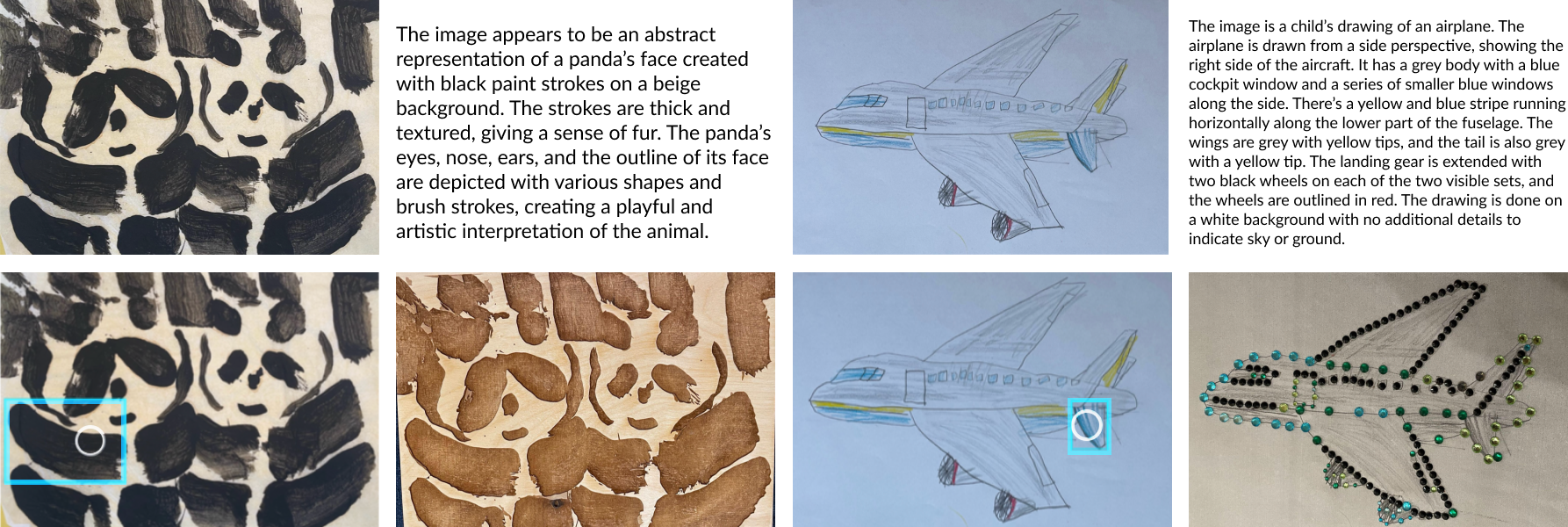}
\Description{This image has 8 panels, arranged in a grid-like format. The four panels on the left are associated with the pandas design probe, and the four on the right are associated with the airplane design probe. Moving from right to left, top to bottom, the panels are: 1) the original painting showing two pandas in black finger paint on white paper; 2) the description of the pandas that says "The image appears to be an abstract representation of a panda’s face created with black paint strokes on a beige background. The strokes are thick and textured, giving a sense of fur. The panda’s eyes, nose, ears, and the outline of its face are depicted with various shapes and brush strokes, creating a playful and artistic interpretation of the animal."; 3) the original airplane drawing, done in pencils on white paper with a blue and yellow strip running across the bottom of the plane; 4) the description of the plane that says "The image is a child’s drawing of an airplane. The airplane is drawn from a side perspective, showing the right side of the aircraft. It has a grey body with a blue cockpit window and a series of smaller blue windows along the side. There’s a yellow and blue stripe running horizontally along the lower part of the fuselage. The wings are grey with yellow tips, and the tail is also grey with a yellow tip. The landing gear is extended with two black wheels on each of the two visible sets, and the wheels are outlined in red. The drawing is done on a white background with no additional details to indicate sky or ground."; 5) a screenshot of the pandas image being explored via touchscreen, a section on the bottom left is highlighted with a blue box indicating where the participant is touching; 6) a laser cut of the pandas image in wood; 7) a screenshot of the airplane image being explored via touchscreen, a section on the bottom right is highlighted with a blue box indicating where the participant is touching; 8) the airplane sketch outlined with raised beads of various colors.}
\caption{Design probes used in Study 2 for in-person participants. (\textit{Left}) The probe associated with a painting of panda bears, and (\textit{right}) the probe associated with a drawing of an airplane. Each cell, in clockwise order from the top left, shows the original child's artwork, an AI-generated description that was read aloud to participants, a tactile representation of the work (laser cut for the pandas, raised dots for the plane), and a screenshot from the Seeing AI touchscreen exploration of the art.}
\vspace{-1em}
\label{DesignProbes}
\end{figure*}

\subsection{AI Explorations with Researcher's Children's Artwork}
See the following examples of artwork provided by researchers run through different AI tools to determine ChatGPT and Be My AI as our final two tools used during our studies. While the results from the Christmas artwork are more mixed, there were some clear dissuading factors for the parent researcher, such as BARD getting the placement of the tree in relation to the animal wrong. These results combined with evaluations of other drawings such as the panda (in which only ChatGPT and Be My AI could recognize pandas on the page) led to us deciding on ChatGPT and Be My AI. Seeing AI was also explored, but as the descriptions were too simplistic, we did not include them in the tables; instead, we report them here. For the panda painting, Seeing AI said: ``A black and white painting''. For the Christmas artwork, Seeing AI said: ``A child's drawing of a Christmas tree and a bird''.

\begin{figure}[h]
\centering
\includegraphics[width=0.5\linewidth]{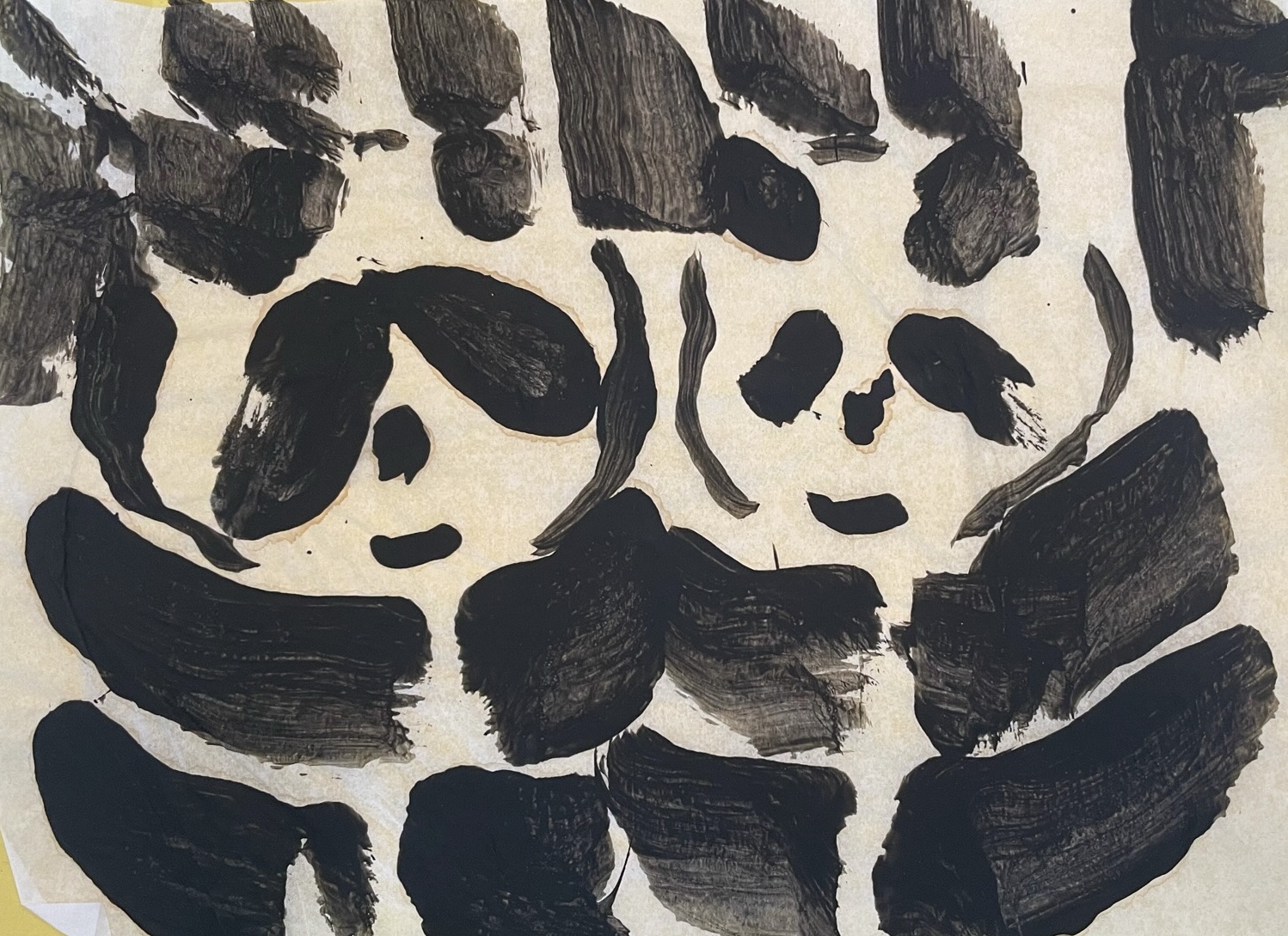}
\Description{A finger painting made with black paint on white paper that has two pandas next to each other. The arms and face of the panda are created via the black finger painting.}
\caption{Sample artwork from a researcher's child---a finger painting of two panda bears in black paint on white paper.}
\label{PandasSample}
\end{figure}

\begin{table*}[t]
  \centering
  \resizebox{\textwidth}{!}
  {
    \begin{tabular}{p{0.3\linewidth} p{0.15\linewidth} p{0.17\linewidth} p{0.25\linewidth} p{0.13\linewidth}}
      \hline
      \textbf{BARD} & \textbf{Bing Chat} & \textbf{LLaVA} & \textbf{ChatGPT} & \textbf{Be My AI} \\
      \hline
      The image you sent me is a close-up of a black and white abstract painting. It features bold black brushstrokes against a crisp white background. The brushstrokes vary in width and thickness, and some have a textured, almost spackled appearance. Here are some additional details about the painting: The brushstrokes appear to have been made with a large brush, and they move in different directions across the canvas. There is a sense of movement and energy in the brushstrokes, and the contrast between the black and white creates a stark and dramatic effect. It is difficult to say for sure what the artist was trying to represent with this painting, as abstract art is open to interpretation. However, the bold brushstrokes could be seen as representing power, strength, or emotion. This type of abstract expressionist painting is often created by applying paint to a canvas in a spontaneous and gestural way. The artist may focus on the physical act of painting and the emotions that it evokes, rather than on creating a representational image.
      & The image you’ve uploaded appears to be an abstract composition of black brush strokes on a white background. The strokes vary in size and shape, creating an intriguing and dynamic pattern. While there is no recognizable image or text, the overall effect is one of artistic expression and spontaneity. It could be interpreted as an abstract painting or a visual representation of movement and energy.
      &  The image you've provided appears to be a piece of abstract art. It features a series of black brush strokes and splatters on a light-colored background. The brush strokes vary in size and shape, creating a dynamic and somewhat chaotic pattern. The art style is reminiscent of ink wash painting, which is a style of East Asian brush painting that uses brush strokes to convey the essence of a scene. However, without more context or a clearer image, it's difficult to provide a more detailed description.
      & This artwork features an abstract panda bear, created with bold and broad strokes of black paint on a beige background. The strokes are textured and vary in shape and direction, giving a sense of spontaneity and movement. The panda's eyes, ears, and the distinctive patches around its eyes are represented in a simplified and somewhat fragmented manner, yet they are still recognizable. The use of negative space plays a crucial role in this composition, as the beige background contributes to defining the form of the panda as much as the black strokes do. The painting exudes a playful and artistic interpretation of the animal, capturing its essence in an abstract, almost calligraphic style. 
      & The image appears to be an abstract representation of a panda's face created with black paint strokes on a beige background. The strokes are thick and textured, giving a sense of fur. The panda's eyes, nose, ears, and the outline of its face are depicted with various shapes and brush strokes, creating a playful and artistic interpretation of the animal. \\
      \hline
    \end{tabular}
  }
  \caption{Results from the Panda Bear artwork (Figure \ref{PandasSample}) run through different AI tools. ChatGPT and Be My AI provided the closest descriptions to how the parent described the artwork.}
\end{table*}

\begin{figure}[h]
\centering
\includegraphics[width=0.5\linewidth]{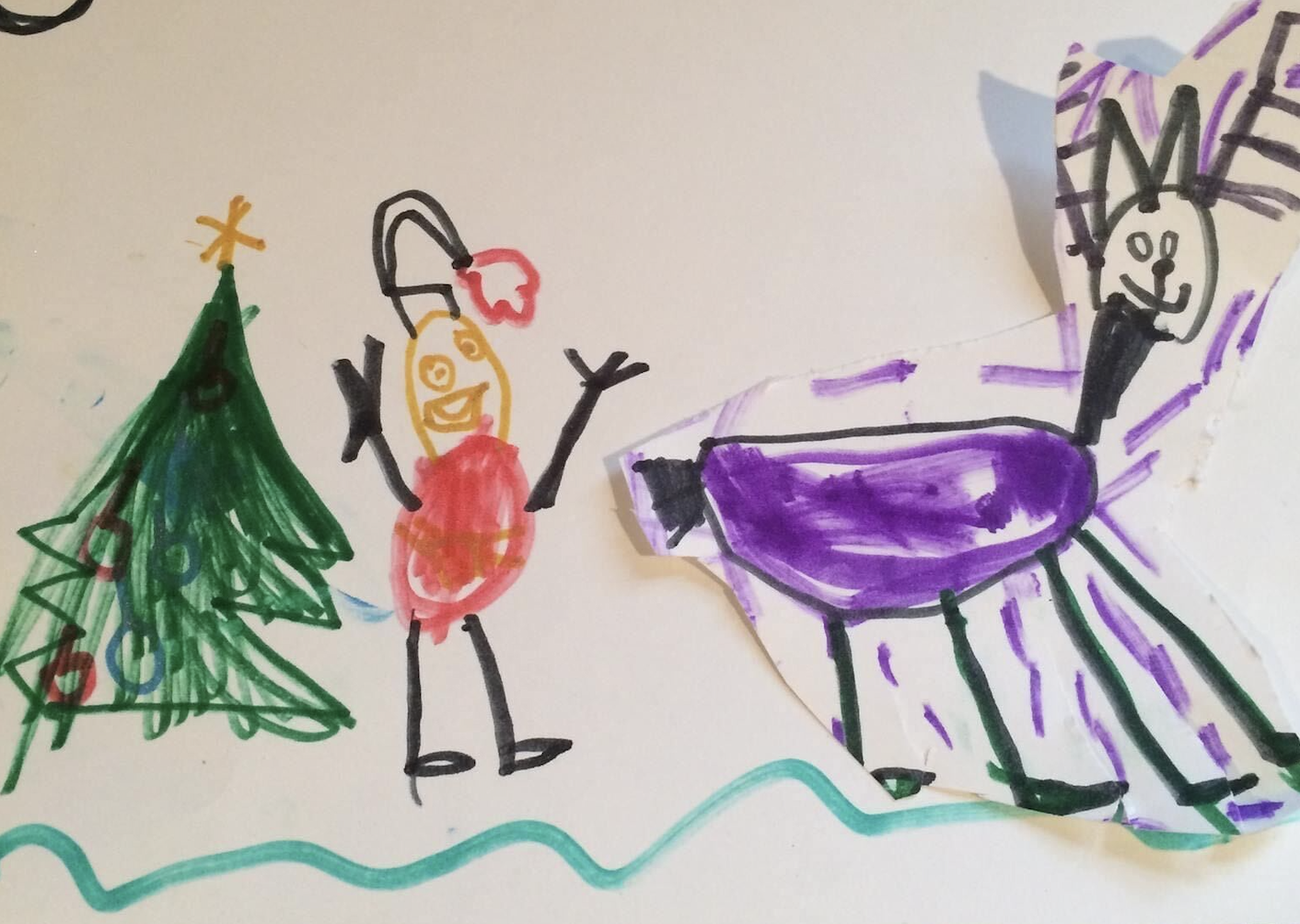}
\Description{A young child's artwork with a green christmas tree on the left, a person with stick figure arms and legs and a red outfit in the middle, and an animal with stick figure legs and a purple body on the right.}
\caption{Sample artwork from a researcher's child containing a drawing of a Christmas tree, a person, and an animal.}
\label{ChristmasSample}
\end{figure}

\begin{table*}[t]
  \centering
  \resizebox{\textwidth}{!}
  {
    \begin{tabular}{p{0.15\linewidth} p{0.25\linewidth} p{0.2\linewidth} p{0.2\linewidth} p{0.2\linewidth}}
      \hline
      \textbf{BARD} & \textbf{Bing Chat} & \textbf{LLaVA} & \textbf{ChatGPT} & \textbf{Be My AI} \\
      \hline
      The image you sent me is indeed a child's drawing of a reindeer and a Christmas tree. The reindeer is drawn in the center of the image. It's brown with a big red nose and antlers made out of two long, pointy triangles. The reindeer has a simple smile and big black eyes. To the right of the reindeer is a Christmas tree. It's a tall green triangle decorated with red circles and yellow squiggles. The top of the tree is decorated with a star. Both the reindeer and the tree are outlined in black crayon.
      & The image you’ve uploaded appears to be a delightful child’s drawing with vibrant colors and playful elements. Let’s break down the details: Christmas Tree: On the left side, there’s a cheerful green Christmas tree adorned with a bright yellow star on top. Squiggly lines suggest ornaments or tinsel decorating the tree. Dancing Figure: In the center, we see a lively figure. It seems to be a person, drawn with enthusiasm. The head is outlined in black, with an orange face and a joyful expression. The body extends downward in red, and the limbs are outstretched as if dancing or celebrating. Small details like eyes, a mouth, and buttons add character to this whimsical creation. Purple Creature: To the right of the dancing figure, there’s a larger purple object. It could be interpreted as an animal or another character. Black lines suggest features like eyes and legs, inviting our imagination to fill in the details. Background: The background is plain white paper, providing a clean canvas for creativity. Blue wavy lines at the bottom might represent water or ground, adding a touch of playfulness. Overall, this artwork exudes childlike wonder and captures the joy of the holiday season.
      &  The image shows a child's drawing on a piece of paper. The drawing features a few simple, colorful figures. On the left, there is a figure that appears to be a person with a smiling face, wearing a red hat and a red coat, standing next to a green tree with a yellow star on top, which could be interpreted as a Christmas tree. To the right, there is a purple figure with a smiling face, which might be a representation of a cat or a similar creature, given the ears and whiskers. The figure is wearing a purple dress or coat. Below the figures, there is a green line that could be interpreted as a path or a river. The drawing is quite abstract and childlike, with a lot of imagination and creativity evident in the simple shapes and bright colors.
      & This is a charming child's drawing. On the left, there’s a green Christmas tree with a star on top and various colored decorations. To the right of the tree stands a joyful character with a round yellow head, wearing a red torso garment and black limbs; they have a festive hat and a broad smile. To the right, there's a zebra with purple stripes instead of the usual black, sporting a friendly face. The background is plain white, and there's a simple wavy blue line, which might represent the ground or a hint of snow. It's a picture full of whimsy and cheer!
      & The picture appears to be a child's drawing featuring three main elements: A Christmas tree on the left, colored green with a yellow star on top and red ornaments. In the center, there is a figure that seems to be a snowman, drawn with a round orange body, black stick arms, a red scarf, and a face with eyes, a nose, and a smile. It has a black top hat with a red band and a small red flower on it. On the right, there is a purple figure that resembles a chair or a throne with black legs and armrests. On the chair, there is a character with a round white face, black stick arms, and legs, and what seems to be purple and black striped ears, possibly representing a rabbit. The background is plain white, and there is a wavy blue line at the bottom, possibly representing the ground or snow.
      \\
      \hline
    \end{tabular}
  }
  \caption{Results from the Christmas artwork (Figure \ref{ChristmasSample}) run through different AI tools.}
\end{table*}

\end{document}